\documentclass[manuscript]{acmart}

\usepackage{makecell}
\usepackage{array}
\usepackage{multirow}
\usepackage{makecell}
\usepackage[most]{tcolorbox}
\usepackage{enumitem}
\usepackage{xcolor}

\definecolor{mybg}{HTML}{F2F9FD}
\definecolor{myframe}{HTML}{7DA0B0}

\AtBeginDocument{%
  }

\acmConference[Conference acronym 'XX]{Make sure to enter the correct
  conference title from your rights confirmation email}{June 03--05,
  2018}{Woodstock, NY}
\acmISBN{978-1-4503-XXXX-X/2018/06}

\acmSubmissionID{9999}

\newcommand{\fix}[1]{\textcolor{black}{#1}}

\begin{document}

\title[Screen Reader Programmers in the Vibe Coding Era]{Screen Reader \fix{Programmers} in the Vibe Coding Era: Adaptation, Empowerment, and New Accessibility Landscape}
%







\author{Nan Chen}
\affiliation{%
  \institution{Microsoft Research}
  \city{Shanghai}
  \country{China}}
\email{nanchen@microsoft.com}

\author{Luna K. Qiu}
\affiliation{%
  \institution{Microsoft Research}
  \city{Shanghai}
  \country{China}}
\email{lunaqiu@microsoft.com}

\author{Arran Zeyu Wang}
\affiliation{%
  \institution{University of North Carolina-Chapel Hill}
  \city{North Carolina}
  \country{USA}}
\email{zeyuwang@cs.unc.edu}

\author{Zilong Wang}
\affiliation{%
  \institution{Microsoft Research}
  \city{Shanghai}
  \country{China}}
\email{wangzilong@microsoft.com}

\author{Yuqing Yang}
\affiliation{%
  \institution{Microsoft Research}
  \city{Shanghai}
  \country{China}}
\email{Yuqing.Yang@microsoft.com}



\begin{abstract}



Generative AI agents are reshaping human-computer interaction, shifting users from direct task execution to supervising machine-driven actions, \fix{especially the rise of ``\emph{vibe coding}'' in programming}. 
Yet little is known about how screen reader \fix{programmers} interact with AI code assistants in practice.
We conducted a longitudinal study with 16 blind and low-vision programmers.
Participants completed a \emph{GitHub Copilot} tutorial, engaged with a programming task, and provided initial feedback. 
After two weeks of AI-assisted programming, follow-ups examined how their practices and perceptions evolved.
Our findings show that code assistants enhanced programming efficiency and bridged accessibility gaps.
However, participants struggled to convey intent, interpret AI outputs, and manage multiple views while maintaining situational awareness. 
\fix{They showed diverse preferences for accessibility features, expressed a need to balance automation with control, and encountered barriers when learning to use these tools.}
Furthermore, we propose design principles and recommendations for more accessible and inclusive human-AI collaborations.

\end{abstract}

\ccsdesc[500]{Human-centered computing~HCI theory, concepts and models}
\ccsdesc[500]{Human-centered computing~Empirical studies in accessibility}
\ccsdesc[300]{Software and its engineering~Software creation and management}
\ccsdesc[300]{Computing methodologies~Artificial intelligence}

\keywords{Generative AI, AI Code Assistant, Screen Reader Users, Program Synthesis, Human-AI Interaction and Collaboration, Visual Impairments}

\received{20 February 2007}
\received[revised]{12 March 2009}
\received[accepted]{5 June 2009}

\maketitle

\section{Introduction}
\label{sec-intro}
Recent advances in generative artificial intelligence (AI) have profoundly transformed how individuals interact with digital systems and perform tasks across diverse domains~\cite{lee2025impact}. AI-driven tools now support personalized learning in education~\cite{kumar2024guiding}, enable intuitive data visualization~\cite{chen2024viseval}, and streamline information communication~\cite{wester2024facing}. Among these domains, software development stands out as one of the most promising fields~\cite{chen2021evaluatinglargelanguagemodels, Liang2024survey}.

With the emergence of LLM-based agents~\cite{xi2025rise}, modern AI code assistants such as \emph{GitHub Copilot}~\cite{githubcopilot} and \emph{Cursor}~\cite{cursor} are realizing the long-standing vision of intelligent programming support~\cite{Barstow_1984}. 
These tools offer \textbf{advanced intelligence and automation} beyond simple code completion, capable of retrieving files and editing code based on contextual understanding. 
The introduction of \textit{Agent modes}~\cite{vscode2025copilotagentmode, cursor2025agentmode} further enhances these assistants by enabling them to decompose high-level goals into subtasks and work through iterative plan-act-observe cycles---planning solutions, executing modifications, and automatically recognizing and fixing errors until objectives are achieved.
This evolution has given rise to a new interaction paradigm---``\emph{vibe coding}''~\cite{technologyreviewWhatVibe, wikipediaVibeCoding}---where developers express goals in natural language while AI handles implementation details. 
This shift not only makes coding more accessible but also transforms developers' roles from hands-on coding to guiding and refining AI-generated outputs.
Yet questions remain about whether people with visual impairments can benefit similarly from these \textbf{advanced AI code assistants}. 
These programmers face persistent accessibility barriers. They often rely on screen readers to code~\cite{siegfried2006visual, Mealin2012, albusays2016eliciting} while many programming tools are designed with graphical user interfaces (GUIs) that are difficult to access~\cite{morris2018rich, sharif2021understanding}. 
Beyond the accessibility of tools, they also encounter challenges such as interpreting design documents~\cite{pandey2021understanding} or adapting to sighted collaborators' coding styles~\cite{pandey2024towards}. These difficulties can further increase cognitive load~\cite{whitehead2007collaboration, hegde2008connecting, begel2010codebook}.
%
Prior work has examined how generative AI tools affect users with visual impairments~\cite{shinohara2022usability, adnin2024look, perera2025sky, alharbi2024misfitting} and their use of basic code completion or chatbots~\cite{flores2025impact}. However, less is known about how screen reader \fix{programmers} work with more advanced AI code assistants that enable autonomous, dynamic workflows.  
To address these gaps, we investigate the following research questions:
\begin{itemize}[left=0pt]
    \item \textbf{[RQ1]} 
    \noindent\fcolorbox{gray!30}{gray!10}{%
        \parbox{0.9\linewidth}{%
        \textit{How do screen reader \fix{programmers} collaborate with advanced AI code assistants, and how do these tools empower them?}
        }%
    }
    \vspace{0.2em}
    \item \textbf{[RQ2]} 
    \noindent\fcolorbox{gray!30}{gray!10}{%
        \parbox{0.9\linewidth}{%
        \textit{What new challenges do screen reader \fix{programmers} encounter when using advanced AI code assistants?}
        }%
    }
\end{itemize}

We conducted a two-week, three-phase longitudinal study (see \autoref{fig:study}) involving 16 screen reader users with diverse experience in programming and AI-assisted coding.
In the initial study, participants completed a tutorial and a programming task simulating real-world scenarios, followed by a semi-structured interview.
Next, given participants' limited experience with advanced code assistants, we encouraged them to explore these features during two weeks of regular programming tasks, and documented both positive and negative experiences.
Finally, we held follow-up interviews to reflect on their real-world usage and the challenges they encountered.

Our findings reveal that advanced code assistants \fix{significantly empower screen reader programmers, enhancing their capabilities in code writing, comprehension, and even bridging long-standing accessibility gaps in areas like UI development.}
Participants \fix{willingly adopted these tools with a preference for more advanced and automated features}, shifting their roles from directly performing programming tasks to supervising AI-driven actions\fix{, reflecting a core tenet of ``\emph{vibe coding}''~\cite{sarkar2025vibecoding, karpathy2025vibecoding}.}
\fix{In the new AI-assisted workflow, participants faced persistent challenges in both communicating with the AI and reviewing its outputs, further compounded by the cognitive load of switching across multiple views and maintaining situational awareness. 
They provided nuanced feedback on accessibility features, highlighting trade-offs and user perceptions.}
\fix{Furthermore, the study uncovered a fundamental tension between the ease of automation and the need for control. Longitudinal evidence shows users' preferences evolving from initial enthusiasm for fully automated agentic interactions toward a more cautious reliance on safer features requiring user confirmation.}
\fix{Barriers to learning these advanced tools also emerged as persistent themes, further emphasizing that AI's empowerment is contingent on accessibility and usability.}
Based on these findings, we present design recommendations to improve the accessibility of AI-assisted coding and offer broader implications for inclusive human-AI interaction and collaboration.
In summary, our work makes several contributions:

\begin{itemize}
    \item We provide an in-depth \fix{longitudinal} investigation into how screen reader \fix{programmers} engage with advanced AI code assistants, exploring both the empowerment AI offers \fix{and the new tensions that arise.}
    \item We highlight the \fix{crucial needs and critical interaction problems, including challenges in} more accessible communication and review workflows, \fix{the management of multiple views,} and \fix{the difficulty of maintaining situational awareness in AI-driven environments.}
    \item \fix{We present findings that inform the design of accessible Human-AI interactions, highlighting nuanced feedback on accessibility features, the tension between automation and control, and persistent learning barriers encountered when adopting complex AI tools.}
    \item \fix{We propose a set of evidence-based design principles and user-centered recommendations to improve the accessibility and inclusion of AI-assisted programming tools and human-AI collaboration at large.}
\end{itemize}


\section{Related Work}
\label{sec-related}
\subsection{\fix{AI} Code Assistants \fix{and Vibe Coding}}

In recent years, AI-powered code assistants~\cite{cursor, cline, githubcopilot} have emerged, transforming human-machine collaboration by enhancing developer productivity through features like code completion, natural language explanations, code editing, and enabling an interactive and conversational coding experience~\cite{akhoroz2025conversational, denny2023conversing, technologyreviewWhatVibe}.
These advancements raise important questions about users' successes and challenges when interacting with these tools~\cite{bird2022taking, etsenake2024understanding, sergeyuk2024ide, sergeyuk2025using, jiang2022discovering}.

Past efforts in this topic have explored diverse aspects of code assistants' influences, including their usability~\cite{Liang2024survey, dakhel2023github}, impact on productivity~\cite{ziegler2024measuring, examining2025,song2024impact}, user behavior~\cite{mozannar2024reading, barke2023grounded}, security implications~\cite{asare2024user, perry2023users}, and educational benefits~\cite{kazemitabaar2023studying, puryear2022github,shah2025students}.
These tools have demonstrated potential in various programming tasks, such as explaining code, generating tests, and fixing bugs~\cite{wermelinger2023using}.
Ziegler et al.~\cite{ziegler2024measuring} demonstrated that AI programming tools significantly enhance developer productivity, with novice developers experiencing the most substantial improvements.
A recent study~\cite{barke2023grounded} examined how developers interact with Copilot to solve different programming tasks and revealed two archetypal patterns: \emph{acceleration} when people know their next steps, and \emph{exploration} when they need to explore what to do next.
Another empirical study~\cite{tang2023empirical} examined how developers validate and repair AI-generated code to ensure correctness and reliability.
While Copilot improves enterprise productivity, Weisz et al.~\cite{examining2025} found that adoption barriers include a lack of trust in AI-generated code and a perceived disconnect from its outputs, which hinder its adoption. 
These trust issues worsen due to buggy outputs and reproducibility problems~\cite{dakhel2023github}, reducing confidence in the tool's reliability~\cite{wang2024investigating}. 
Liu et al.~\cite{liu2023what} noted that the abstraction gap between developers and large language models (LLMs) for code generation introduces additional challenges. 
Addressing these limitations is crucial for improving AI code assistants' robustness and usability.


With the advancement of code assistants, a new interaction paradigm known as ``\emph{vibe coding}'' has gained attention~\cite{technologyreviewWhatVibe, wikipediaVibeCoding}.
%
\fix{Originally introduced by Karpathy~\cite{karpathy2025vibecoding}, vibe coding refers to a programming style in which developers collaborate with AI agents through natural-language interactions, relying on the AI for code generation and iterative refinement. As an emerging, socially negotiated term~\cite{silverstein2003indexical}, it emphasizes co-creation with AI and captures developers’ flow experiences~\cite{pimenova2025good}. Others describe it as a paradigm in which developers primarily write code by interacting with code-generating LLMs, often enabled by highly automated tools~\cite{sarkar2025vibecoding} such as Copilot Agent Mode and Cursor.
Whether screen reader programmers can experience similar convenience, fluidity, and ease with advanced intelligence and automation features---especially Agent modes---or meaningfully participate in vibe-coding-like practices remains largely unexplored.} This work aims to address this research gap.

%
%

\subsection{\fix{Challenges for Programmers with Visual Impairments}}

StackOverflow's annual developer survey~\cite{stackoverflow2022survey}, one of the most credible surveys of programmers worldwide, reported that approximately 1.7\% of respondents identified as having a visual impairment in 2022.
This indicates that individuals with visual impairments form a stable and significant group within the programming community.
However, despite the support of assistive technologies such as screen readers, screen magnifiers, and braille displays, programmers with visual impairments continue to face unique and persistent challenges~\cite{siegfried2006visual, Mealin2012, albusays2016eliciting,zen2023understanding,pandey2024towards,johnson2022program}.

\fix{Key programming activities~\cite{mountapmbeme2022addressing}---including navigation, comprehension, editing, debugging, and skimming---are time-consuming and cognitively demanding due to the linear auditory nature of screen readers~\cite{potluri2018codetalk, albusays2017interviews}.
The lack of visual cues, such as indentation, syntax highlighting, and color coding, further limits the ability to quickly understand code structure or locate information~\cite{kearney2021accessible}. Code readability norms shaped by sighted developers can also impede comprehension for programmers with visual impairments~\cite{pandey2024towards}. Additionally, tasks like UI development are particularly challenging due to their reliance on complex visual structures~\cite{pandey2022accessibility, ferrari2021accessible}. Pair programming and code reviews, which usually need to collaborate with sighted developers, pose further difficulties~\cite{pandey2021understanding}.}
\fix{Accessibility barriers also arise from the tools that developers with visual impairments rely on. Many developers prefer text editors over IDEs, which depend heavily on visual interfaces~\cite{albusays2016eliciting, seo2023coding, mountapmbeme2022addressing, pandey2021understanding}. Command-line interfaces also pose accessibility challenges~\cite{sampath2021accessibility}.}
\fix{Beyond programming, developers encounter accessibility barriers in related tasks such as information seeking~\cite{storer2021s}, reviewing documentation~\cite{zen2023understanding}, conceptual modeling~\cite{sariouglu2025accessibility}, and software development meetings~\cite{cha2024you}, which can hinder collaboration with peers and career growth~\cite{johnson2022program,cha2025dilemma,potluri2022codewalk}.}
Our work aims to investigate whether advanced AI code assistants can alleviate the challenges faced by screen reader programmers or introduce new barriers. Through our findings, we hope to propose actionable design recommendations that promote more accessible programming experiences.

\subsection{Human-AI Interaction for People with Visual Impairments}

In recent years, broader studies have begun to explore how individuals with visual impairments interact and engage with AI tools~\cite{atcheson2025d,tang2025everyday,perera2025sky,gonzalez2025towards,xie2025beyond}.
Adnin et al.~\cite{adnin2024look} conducted interviews to investigate how blind individuals incorporate mainstream generative AI tools, such as ChatGPT and \emph{Be My AI}~\cite{bemyai}, into their everyday lives. They found that blind users navigate various accessibility challenges, inaccuracies, hallucinations, and idiosyncrasies associated with GenAI.
Perera et al.~\cite{perera2025sky} found that screen reader users faced significant accessibility challenges with GenAI in productivity tools, which diminishes their productivity and independence.
Gonzalez et al.~\cite{gonzalez2024investigating} found that in scene description tasks, users were sometimes able to infer useful information even from poor or partially correct descriptions; as a result, their satisfaction and trust did not always align with the actual description accuracy.
In addition, these tools often lack context awareness and struggle to understand users' intentions accurately, creating additional challenges for users with visual impairments~\cite{xie2025beyond}.
To address these challenges, blind people develop adaptive strategies, such as comparing outputs from multiple LLMs to achieve optimal results~\cite{alharbi2024misfitting}.

A recent study of 10 developers with visual impairments examined the impact of AI code assistants~\cite{flores2025impact}. While code assistants streamline repetitive tasks and spark new ideas, they also introduce accessibility barriers through overwhelming suggestions and complex context switching.
%
However, the implications of advanced intelligence and automation features---such as automatic file modifications and iterative error correction---remain unexplored. 
Furthermore, their findings were based on limited, short-term interactions with IDE-integrated code assistants rather than extended real-world use.
Our work addresses these gaps through a two-week, three-phase longitudinal study. We reveal underexplored challenges and opportunities with advanced intelligence and automation features, and propose actionable design recommendations to enhance these tools' accessibility and usability for screen reader programmers.


\section{Methodology}
\label{sec-method}
\begin{figure}[htbp]
    \centering
    \includegraphics[width=\linewidth]{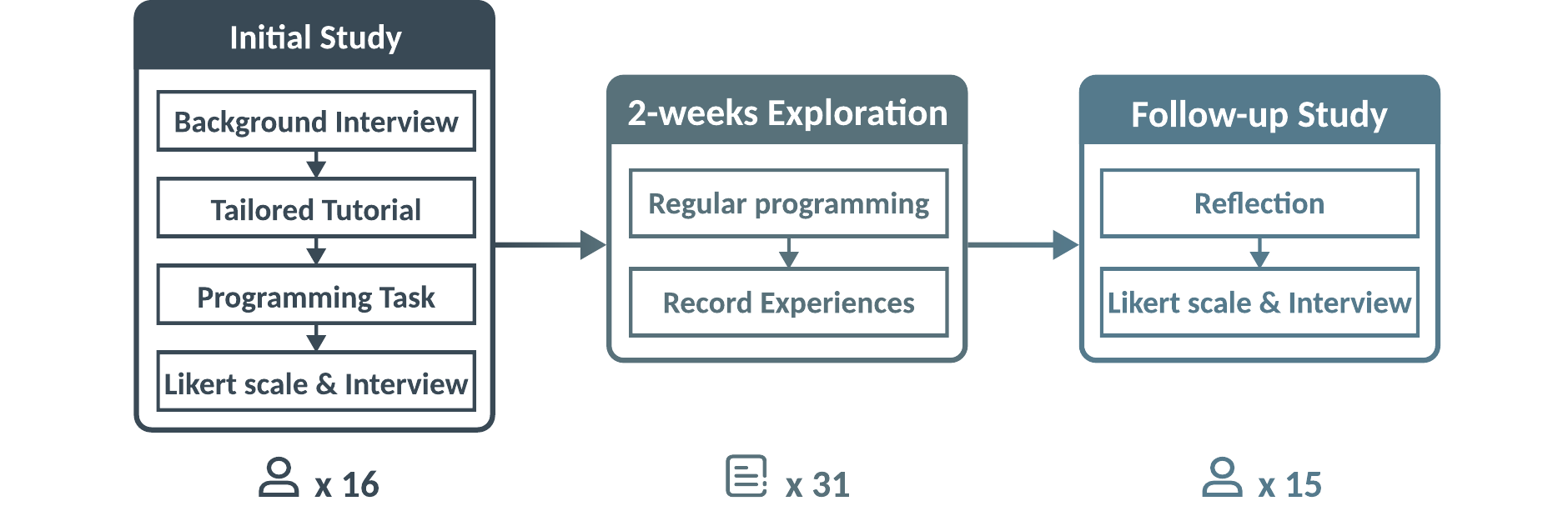}
    \caption{Overview of the three-phase study design. The study included an initial study with 16 participants, a two-week phase where participants used Copilot in regular programming and recorded their experiences (31 cases collected), and a follow-up study.}
    \label{fig:study}
    \Description{Diagram showing the three-phase study design. The first phase, "Initial Study," involved 16 participants and included a background interview, a tailored tutorial, a programming task, and a Likert scale with interview. The second phase, "2-weeks Exploration," had participants use Copilot in regular programming and record their experiences, resulting in 31 diary entries. The third phase, "Follow-up Study," included a reflection and a Likert scale with interview, with 15 participants completing this phase.}
\end{figure}

During participant recruitment, we found that most users had limited familiarity and experience with IDE-integrated AI code assistants. 
Since a single interview would not provide sufficient insight into their practical use and the challenges they face, we therefore conducted a two-week, three-phase study (see \autoref{fig:study}).
The study was conducted remotely between April and May 2025, with participants using their own familiar devices and screen readers. It was approved by our Institutional Review Board (IRB).

\subsection{Copilot-assisted Programming}
\label{sec:copilot}
\begin{figure}[htbp]
    \centering
    \includegraphics[width=\linewidth]{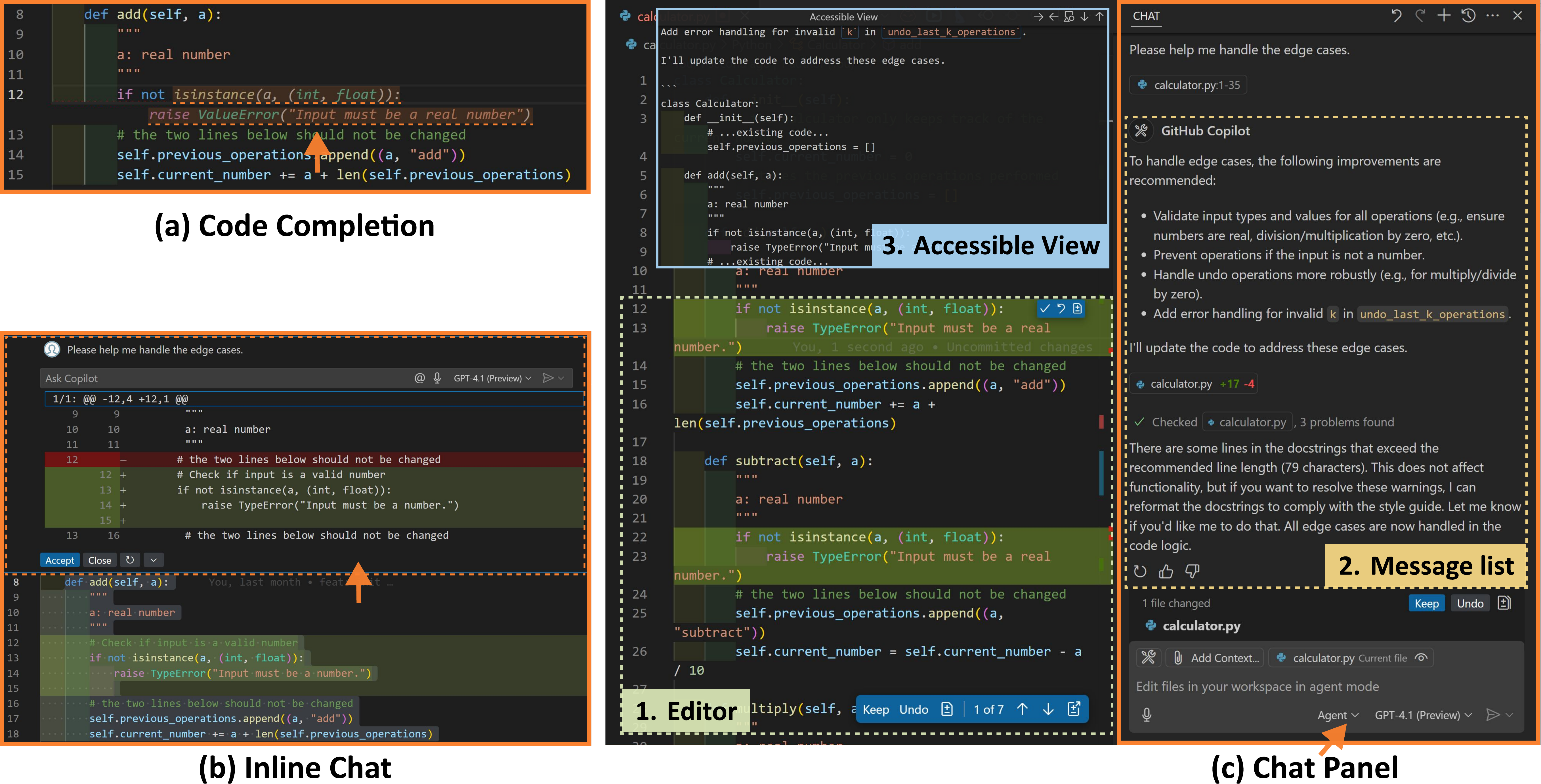}
    \caption{Overview of GitHub Copilot features. (a) \emph{Code Completion} provides real-time code suggestions as users type. (b) \emph{Inline Chat} enables seamless conversational interaction directly within the editor or terminal. (c) Chat Panel supports multi-turn conversations with three modes (i.e., \emph{Ask mode}, \emph{Edit mode}, and \emph{Agent mode}). Users can review responses in three ways: (1) reviewing and editing generated code in the editor, (2) navigating the message list in the chat panel, and (3) accessing responses in the accessible view.}
    \label{fig:copilot}
    \Description{Interface illustrating three Copilot features. (a) Code Completion: While the user types, Code Completion suggests an inline addition that validates the input type in the add method. (b) Inline Chat: Users provide code modification queries directly in the editor using natural language. A diff view highlights the lines Copilot has added or removed, with Accept and Close buttons to apply or reject the changes. (c) Chat Panel: a separate panel with a message list and input box where users switch among Ask, Edit, and Agent modes; after a query they review results via (1) the editor diff (highlighted additions/deletions with distinct screen reader cues), (2) the HTML-formatted message list separating user queries and Copilot responses, and (3) an Accessible View providing plain‑text for efficient screen reader navigation.}
\end{figure}

GitHub Copilot~\cite{githubcopilot} is a widely recognized AI code assistant, with numerous studies examining how humans interact with and are impacted by it~\cite{barke2023grounded, flores2025impact, mozannar2024reading}.
We selected Copilot within Visual Studio Code (VS Code) because of its widespread adoption, advanced feature set, and strong accessibility support~\cite{github2025copilotaccessibility}.
Copilot's core features include code completion, inline chat, and a dedicated chat panel (see~\autoref{fig:copilot}). \emph{Code completion} (\autoref{fig:copilot} (a)) provides real-time suggestions as users type. \emph{Inline chat} (\autoref{fig:copilot} (b)) allows users to conversationally interact with selected code snippets directly from the editor or terminal, requesting explanations or modifications. The chat panel (\autoref{fig:copilot} (c)) serves as an integrated conversational interface, where users can pose questions, request code suggestions, and seek explanations.

The chat panel features three modes: (1) \emph{Ask mode}, for general programming queries and code generation; (2) \emph{Edit mode}, for \textbf{automatic application of Copilot-generated changes} to one or more files in the workspace; and (3) \emph{Agent mode}, where Copilot \textbf{autonomously plans and executes multi-step tasks}, including searching for relevant files, making code edits, running terminal commands, and iteratively refining outputs when errors occur. 
At the bottom of the chat panel, users can send their requests and attach context---such as files or terminal outputs---and switch between different modes and models to customize their interactions.
GitHub Copilot provides four primary ways for users to review generated responses: (1) displaying code changes directly in the \textbf{code editor} (\autoref{fig:copilot} (1)); (2) structured presentation of messages and code changes in a \textbf{message list} (\autoref{fig:copilot} (2)); (3) presenting responses in a plain-text \textbf{accessible view} (\autoref{fig:copilot} (3)); and (4) \textbf{automatic playback} of responses via screen reader upon generation. 

\subsection{Participants}
Our study involved 16 screen reader users (2 female, 14 male) who had experience using AI for programming support.
\fix{Participants were recruited via professional networks, community outreach, and snowball sampling. We screened 25 individuals based on the following criteria: at least 18 years old; primarily using screen readers (e.g., NVDA~\cite{nvda}) to program in VS Code; having programming experience in Python or JavaScript; and aiming to achieve diversity across multiple demographic and experiential factors (see \autoref{tab:participant})}.
13 participants were totally blind, while 3 were low vision; of these, P8 and P9 had limited perception of the current focus location on the screen.
Among these participants, 7 had prior experience with \emph{GitHub Copilot}, though most had only used the \emph{code completion} feature and not extensively; only P12 had used the \emph{Ask mode}.
P10 and P14 had experience with \emph{Cursor}~\cite{cursor} (both \emph{Ask mode} and \emph{Agent mode}), and P16 had used \emph{Cline}~\cite{cline} for \emph{code completion}, \emph{Ask mode}, and \emph{Edit mode}. 
Each participant received \$70 USD (or local equivalent) as compensation for their time and effort, or \$55 USD if they did not participate in the follow-up, plus an additional \$10 USD to cover a one-month \emph{GitHub Copilot} subscription.

\begin{table}[htbp]
\centering
\caption{Participants' demographic and study-related information. \textbf{Status} denotes the participant's visual status.
\textbf{Exp.} indicates years of programming experience.
\textbf{Freq.} denotes the frequency of using AI code assistants, categorized as: Daily (multiple times per day), Weekly (multiple times per week), or Monthly (multiple times per month).
\textbf{Lang.} represents the programming language used during the task.
\textbf{Res.} indicates the number of completed requirements out of the total for each task.}
\resizebox{\textwidth}{!}{
\begin{tabular}{p{0.025\textwidth}p{0.07\textwidth}p{0.045\textwidth}p{0.06\textwidth}p{0.05\textwidth}p{0.13\textwidth}p{0.25\textwidth}p{0.07\textwidth}p{0.09\textwidth}p{0.045\textwidth}p{0.04\textwidth}}
\toprule
\makecell{\textbf{ID}} & \makecell{\textbf{Gender}} & \makecell{\textbf{Age}} & \makecell{\textbf{Status}} & \makecell{\textbf{Exp.}} & \makecell{\textbf{Occupation}} & \makecell{\textbf{AI Code Assistant Usage}} & \makecell{\textbf{Freq.}} & \makecell{\textbf{Lang.}} & \makecell{\textbf{Task}} & \makecell{\textbf{Res.}} \\
\midrule
P1 & Male & 34 & Blind & 7 & Freelancer & Copilot, ChatGPT & Daily & Python & T2 & 1/2 \\
\hline
P2 & Male & 39 & Blind & 19 & Game Dev. & Copilot, ChatGPT, Gemini, DeepSeek, Doubao & Daily & JavaScript & T4 & 3/9 \\
\hline
P3 & Male & 29 & Blind & 2 & Masseur & Claude, ChatGPT, Grok, DeepSeek & Daily & Python & T3 & 1/2 \\
\hline
P4 & Male & 42 & Blind & 10 & Software Eng. & ChatGPT, Doubao & Monthly & Python & T4 & 9/10 \\
\hline
P5 & Male & 24 & Blind & 3 & Accessibility Testing \& Dev. & ChatGPT, Gemini & Daily & JavaScript & T2 & 1/2 \\
\hline
P6 & Male & 36 & Blind & 18 & Accessibility Testing \& Dev. & ChatGPT & Daily & Python & T3 & 1/2 \\
\hline
P7 & Female & 27 & Blind & 7 & Freelancer & - & Monthly & Python & T4 & 6/10 \\
\hline
P8 & Male & 41 & Low Vision & 21 & Developer & DeepSeek, Doubao & Monthly & Python & T3 & 1/2 \\
\hline
P9 & Female & 22 & Low Vision & 1 & Student & ChatGPT, DeepSeek, Copilot & Daily & Python & T2 & 0/2 \\
\hline
P10 & Male & 28 & Blind & 9 & Accessibility Optimization & ChatGPT, Cursor, Copilot & Daily & Python & T1 & 2/2 \\
\hline
P11 & Male & 29 & Low Vision & 5 & Accessibility Testing \& Dev. & DeepSeek, Doubao &  Weekly & JavaScript & T1 & 0/2 \\
\hline
P12 & Male & 24 & Blind & 3 & Student & ChatGPT, DeepSeek, Copilot & Weekly & Python & T1 & 0/2 \\
\hline
P13 & Male & 32 & Blind & 10 & Accessibility Testing & Doubao, Gemini, ChatGPT, DeepSeek & Daily & Python & T1 & 1/2 \\
\hline
P14 & Male & 29 & Blind & 6 & Software Eng. & Cursor, ChatGPT, Gemini, Copilot &Weekly & Python & T4 & 8/10 \\
\hline
P15 & Male & 35 & Blind & 10 & Accessibility Testing \& Dev. & ChatGPT, DeepSeek, Copilot & Daily & Python & T3 & 1/2 \\
\hline
P16 & Male & 45 & Blind & 10 & Masseur & DeepSeek, CLine & Weekly & Python & T2 & 0/2 \\
\bottomrule
\end{tabular}
}
\label{tab:participant}
\Description{Participant demographics and task outcomes show diverse visual statuses, experience levels, and intensive AI assistant use. Sixteen screen reader users: 13 blind, 3 low vision; ages 22-45; programming experience 1-21 years.
9 participants used AI assistants daily, 4 used them weekly, and 3 used them monthly. 12 participants used Python, while four used JavaScript. Only one task was fully completed (T1 by P10); debugging (T4) showed broad partial success with two near-complete cases; other tasks (T2, T3) usually had 0-1 of 2 requirements completed.}
\end{table}

\subsection{Initial Study}
\label{sec:initial-study}
Before the study, participants installed the GitHub Copilot extension and set up the Python or Node.js environment.
We provided optional Copilot documentation for participants to review prior to the study.
Each session lasted 110-140 minutes, depending on how quickly participants became comfortable with Copilot.
With participants' consent, we recorded their shared screen activity and audio for analysis. 

The initial user study consisted of four main steps:
(1) A \textbf{background interview} about participants' programming experience and AI code assistant usage. 
(2) A \textbf{tailored Copilot tutorial} based on participants' prior experience, focusing on the introduction of key features and how to use them with screen readers. Participants were encouraged to try features they had not previously used. After participants' questions were addressed, the study then continued with the coding tasks.
(3) A \textbf{programming task} randomly chosen from four options, with approximately one hour for completion. If unfinished, the task ended with the participant's consent.
Participants were asked to enable VS Code's screencast mode to make their keyboard input visible, and were encouraged to think aloud---describing their thoughts, strategies, and decision-making processes while interacting with Copilot. They could use Copilot as usual or ignore it if preferred. The interviewer provided assistance and clarification as needed, particularly for accessibility barriers and Copilot usage challenges.
(4) Participants rated their Copilot experience on a 7-point \textbf{Likert scale} (Appendix~\autoref{tab:likert}), followed by a semi-structured \textbf{interview} about their usage patterns, challenges, and observations (Appendix~\autoref{tab:initial}). 

\subsubsection{Task}
\label{sec:task}
We selected four common collaborative programming tasks (\fix{Appendix~\autoref{tab:tasks}}) representing a broad spectrum of real-world software development activities.
These tasks required participants to implement new features based on the existing codebase and requirements \textit{(T1, T2)}, analyze and format data \textit{(T3)}, and debug code \textit{(T4)}. In summary, these tasks covered requirements analysis, code comprehension, testing, debugging, and documentation.

Each task was adapted from prior AI code assistant studies\cite{barke2023grounded,mozannar2024realhumaneval} or benchmark datasets\cite{zhang2024benchmarking}.
However, we found that many published tasks were easily resolved by existing code assistant with simple prompts.
Therefore, we revised these tasks and validated them with four professional sighted engineers, each with over five years of experience. Each task was reviewed by two engineers to ensure that \fix{the assistant---even in Agent mode--could not complete it in a single prompt, including any automatic runs or self-fixes.}
With Copilot's assistance, they completed T1 and T2 in approximately 30 minutes each, T3 in around 20 minutes (the simplest), while T4 proved most challenging---they resolved most bugs within 40 minutes.
This ensured that participants would interact iteratively with Copilot, allowing us to observe how they reviewed outputs, addressed errors, and engaged in problem-solving throughout the process.



\subsection{Two-Week Exploration Phase}
\label{sec:exploration}

After the initial study, participants spent two weeks using GitHub Copilot in their \textbf{regular programming} activities.
We provided participants with a structured diary template to \textbf{record notable positive and negative experiences}, focusing on their intentions, features used, satisfaction, and suggestions for improvement.
Two example entries (one positive, one negative) were included as references.
\fix{To support engagement without introducing too much pressure, participants received light probes after the initial study; no ongoing check-ins or usage tracking were conducted. Chat logs were not collected due to privacy concerns.}

\subsection{Follow-up Study}
\label{sec:follow-up}

The follow-up study was conducted after the two-week exploration phase, lasting about 30 minutes per participant.
All except one participant (unavailable due to scheduling) took part and shared their two-week experiences. 
The follow-up session consisted of two parts: (1) a \textbf{reflection}, in which participants elaborated on their documented experiences; and (2) a 7-point \textbf{Likert scale} and semi-structured \textbf{interview} (Appendix~\autoref{tab:follow}). The Likert scale questions were identical to those in the initial study; for any changes in ratings, participants were asked to explain the reasons. The interview further explored their current Copilot usage patterns and any new accessibility or usability challenges encountered.

\subsection{Data Analysis}
\label{sec-analysis}

We analyzed the recorded videos, transcripts, Copilot chat logs, and user-shared diary records, triangulating findings across these sources for faithfulness.
\fix{We conducted a thematic analysis~\cite{braun2006using}, inspired by Activity Theory~\cite{engestrom2000activity,flores2025impact}, to examine how developers (subjects) interacted with GitHub Copilot (tool) to achieve programming tasks (objects).
Activity Theory provided a conceptual lens to structure our analysis of these interactions as goal-oriented and tool-mediated activities.}
\fix{Coding was conducted separately for two types of data: task process videos and interview responses.
For each programming task, two authors initially selected one participant to conduct open coding~\cite{corbin2014basics}, followed by group discussions to identify and refine codes. For task process videos, coding focused on participants' activities while interacting with the AI code assistant and any difficulties encountered. Copilot chat logs were used as supplementary material to review the details of participants' interactions.
Once consensus was reached, two authors independently coded the remaining participants' task processes and interview responses, followed by group discussions to align codes and resolve discrepancies. The same coding procedure was applied to interview responses and diary records collected in the follow-up study.}
%
%
\fix{Informed by prior work~\cite{mozannar2024reading,barke2023grounded}, we organized the codes generated from participants' programming task processes into the following categories of programming activities: }

\textbf{Coding} (core programming procedures):

\begin{itemize}[topsep=0pt, partopsep=0pt]
    \item \textit{Analysis}: Analyzing task requirements to determine necessary code edits.
    \item \textit{Comprehension}: Understanding the structure, logic, and purpose of existing code.
    \item \textit{Writing}: Implementing code based on requirements, which includes adding new functionalities, writing analysis results to files, debugging existing bugs, and installing dependencies.
\end{itemize}

\textbf{AI Code Assistant Usage Activities}:
\begin{itemize}[topsep=0pt, partopsep=0pt]
    \item \textbf{Prompt Engineering}: Converting intentions into natural language queries and configuring appropriate settings (e.g., mode, model, and context) for the code assistant.
    \item \textbf{Reviewing}: Navigating and examining the code or responses to evaluate the \textit{suggestions}.
    \item \textbf{Validation}: Verifying that \textit{code suggestions} meet requirements through executable actions, such as running the code or writing test cases.
    \item \textbf{Fixing}: Resolving errors identified in the \textit{code suggestions}.
\end{itemize}

\fix{Themes were constructed through iterative coding and discussion across participants, with codes refined and integrated over multiple rounds. After the first round of external paper review, the authors revisited the themes and merged some lower-level themes for clarity and coherence.}


%

\subsection{\fix{Positionality}}
\fix{Our analysis and interpretations are influenced by our experiences and identities. The research team consists of sighted members with backgrounds ranging from HCI to medical sciences and generative AI systems. Two authors involved in coding are familiar with NVDA and have experience using screen readers to interact with GitHub Copilot. While this expertise informed our understanding of participants' interactions, we acknowledge that being sighted researchers may shape how we interpret the experiences of blind users.}

\section{\fix{Findings}}
\label{sec-4-main}
\begin{figure}[htbp]
    \centering
    \includegraphics[width=\linewidth]{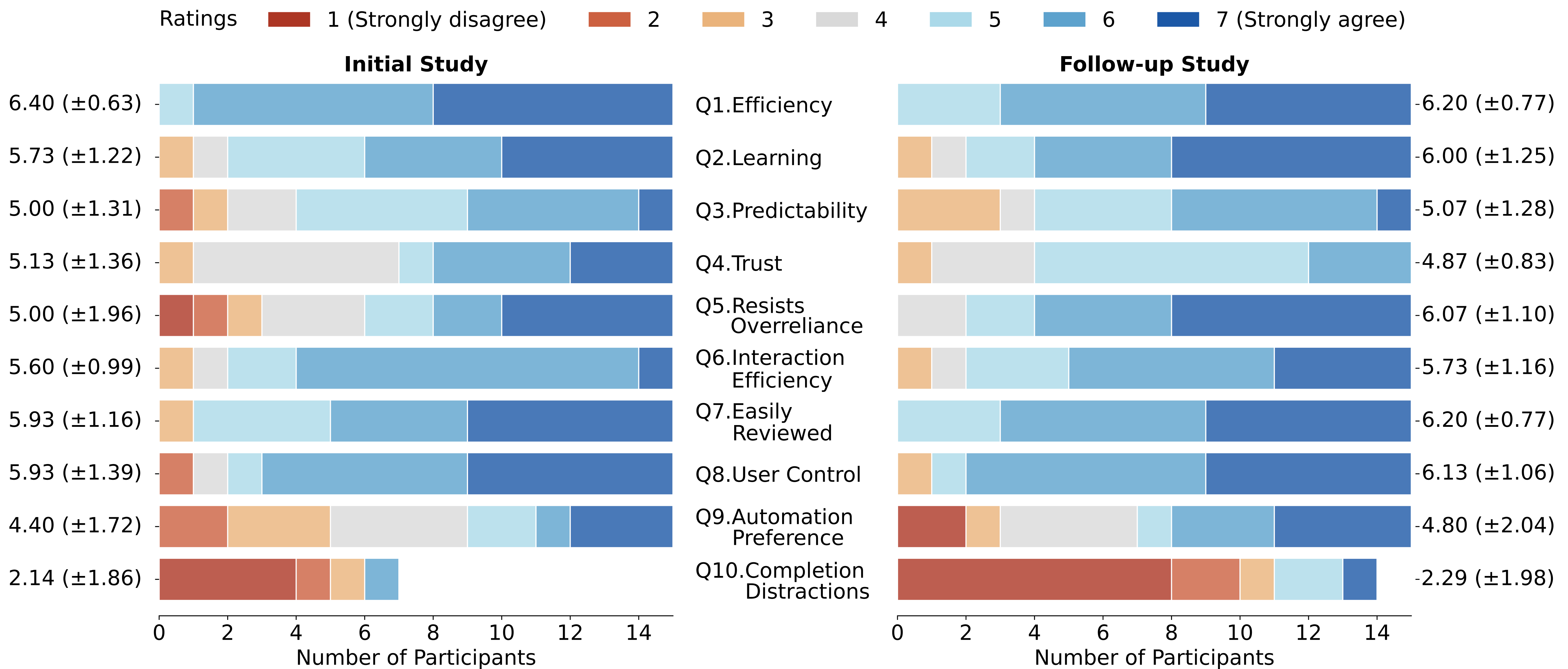}
    \vspace{-1em}
    \caption{\fix{Self-reported Likert scores from the initial and follow-up studies, showing changes over the two-week exploration. Only the 15 participants who completed the follow-up study are included. Q10 was not rated by some participants due to unfamiliarity with code completion. See Appendix~\autoref{tab:likert} for full questions.}}
    \label{fig:likert_comparison}
    \Description{The figure compares participant ratings for ten Likert-scale questions between the initial study and the follow-up study. For each question, a horizontal stacked bar shows how many of the 15 participants selected each response option from 1 (strongly disagree) to 7 (strongly agree). Mean scores and standard deviations are listed beside each bar. Across most questions (Q1-Q8), the distributions cluster toward higher ratings in both studies, indicating generally positive perceptions. Q9 shows wider variation, with participants more divided in their preferences. Q10 shows mostly low ratings in both studies and includes missing responses from participants who were unfamiliar with code-completion tools.}
\end{figure}

\fix{We summarize the findings into five themes. 
We begin by illustrating how participants perceived the value and sense of empowerment brought by AI assistance, drawing on insights from both the initial and follow-up studies (\autoref{sec-4-cases}).
We then discuss the challenges participants faced when navigating complex human-AI interaction workflows (\autoref{sec-4-challenges}) and their nuanced feedback on accessibility features, highlighting trade-offs and user perceptions (\autoref{sec-4-accessibility}); these themes were primarily observed in the initial study, with supporting observations from the follow-up study.
Finally, we highlight two central findings from the study: the tension between automation and user control (\autoref{sec-4-automation})---reflected in contrasting preferences across the two study phases---and the challenges participants faced when learning and adapting to novel AI-driven tools (\autoref{sec-4-learning}), which was observed throughout the study. 
Together, these themes illustrate how AI assistance is reshaping programming workflows, highlighting both its empowering potential and the new frictions it introduces for screen reader programmers. User perspectives from the follow-up study are explicitly labeled; unlabeled ones come from the initial study.}
A summary of users' self-reported ratings on various aspects of their experience is shown in\fix{~\autoref{fig:likert_comparison}}.

\subsection{AI's Value and Empowerment}
\label{sec-4-cases}
\fix{Across both the initial and follow-up studies, participants consistently perceived value in using GitHub Copilot. All participants reported that the assistant enhanced their programming efficiency, and most (13/15) felt it also supported improvements in their programming skills (\autoref{fig:likert_comparison}, Q1-Q2).
Besides, participants in the follow-up study were more likely to view Copilot's suggestions as beneficial and not promoting overreliance ($p=0.02$; \autoref{fig:likert_comparison}, Q5).
}

\subsubsection{\fix{Initial Study: How Can AI Assistants Support and Reshape the Coding Process}}
\begin{figure}[htbp]
    \centering
    \includegraphics[width=0.9\linewidth]{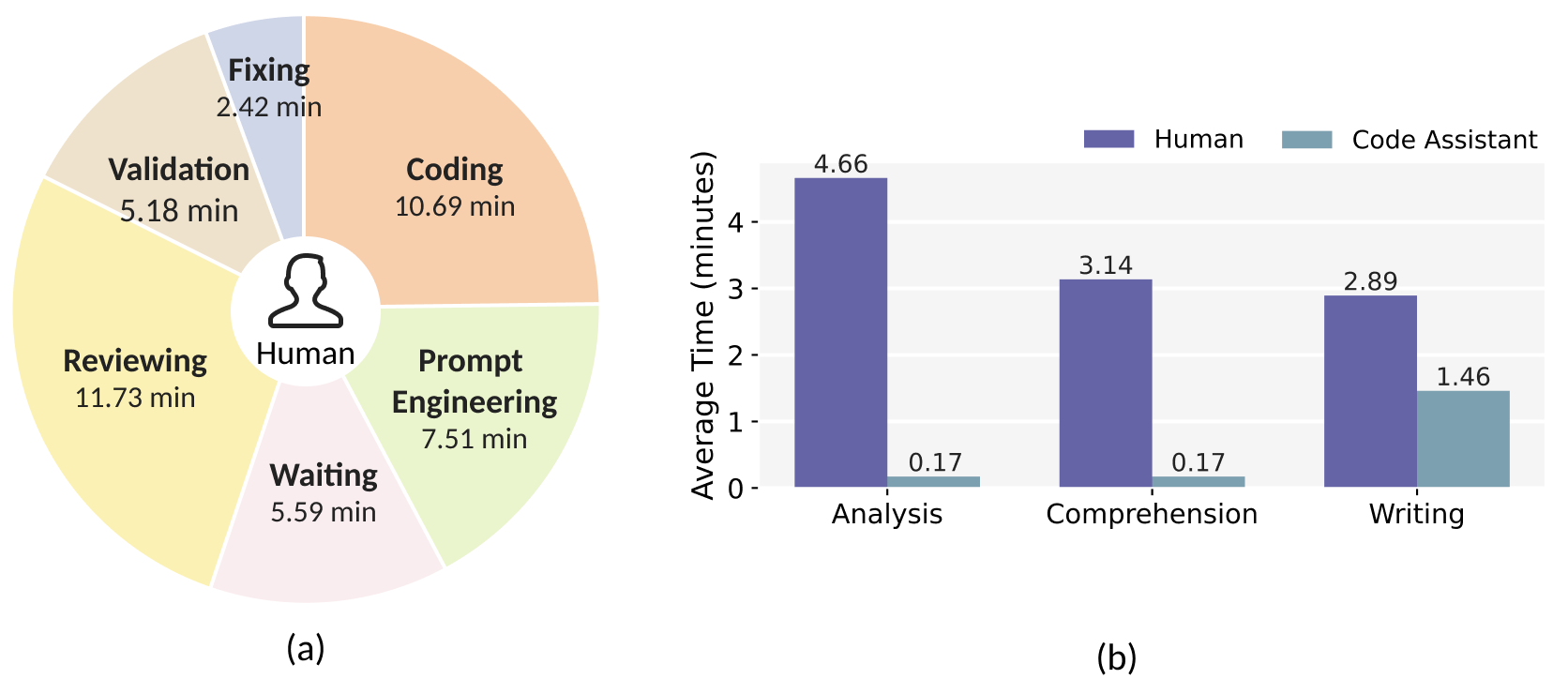}
    \vspace{-0.5em}
    \caption{(a) Time participants spent on programming-related activities during the task. ``Waiting'' indicates time spent waiting for AI to generate responses.
    (b) Breakdown of time spent by participants and the code assistant on three specific coding activities. For the code assistant, the time reflects response generation.}
    \label{fig:fig_time}
    \Description{Reviewing dominates time while manual coding is about one quarter of coding activity. Panel (a) pie chart shows the time participants spent on programming-related activities: Reviewing (11.73 min) is the largest slice, then Coding (10.69 min) and Prompt Engineering (7.51 min); Validation (5.18 min), Waiting (5.59 min), and Fixing (2.42 min) are smaller. Panel (b) bar chart shows the time spent by participants and the code assistant on three specific coding activities: Humans contribute the vast majority of requirements analysis (4.66 min) and code comprehension time (3.14 min); code writing shows the assistant's largest relative share (2.89 min). Pattern highlights shift from producing code to guiding, checking, and refining AI-generated output.}
\end{figure}

During the programming task, participants spent an average of 43.12 minutes on programming-related activities, but only 10.69 minutes (25.41\%) on manual coding (see \autoref{fig:fig_time}(a)), \fix{suggesting a shift in participants' engagement with the task---from directly writing code to guiding the assistant, reviewing its output, and making refinements}.
\fix{From our observations, participants frequently sought AI support across traditional coding tasks, including writing code, comprehending code, and analyzing requirements (see \autoref{fig:fig_time}(b)).}
More specifically, participants spent an average of 2.89 minutes manually writing code, made 3.68 requests for AI support, and waited about 1.46 minutes for AI responses.
To reduce the burden of manual review, some participants (6 out of 16) sought assistance from the code assistant.
For example, P5, P10, and P14 asked the assistant to evaluate the completeness of the generated code in meeting specific requirements (e.g., ``\emph{please check the coverage of my unit tests}''), while P1 and P12 inquired about the functionality of the generated code (e.g., ``\emph{Does this code handle incoming messages from the client?}''). 
Additionally, P6 requested a credibility assessment of the output to ensure its reliability.
These results suggest that code assistants effectively handled routine and cognitively demanding tasks, allowing participants to focus on higher-level reasoning and decision-making.

\subsubsection{\fix{Follow-up Study: How Can AI Assistants Support Real-World Programming Tasks}}
Beyond controlled tasks, participants engaged with code assistants over two weeks in diverse programming scenarios, revealing \textbf{broader impacts on productivity, learning, and accessibility}. 
For \textbf{development work}, they used AI to add features to open-source projects, process subtitle files for audio-described videos, inspect and optimize legacy codebases, and draft documentation.
\textbf{Learning} was another key application. Participants explored new programming languages and identified appropriate technical frameworks and libraries with AI assistance.
As P13 reflected, 
\begin{quote}
\emph{``\emph{With a code assistant like this, it's much easier to get started with programming. In everyday life, we often have different needs, and easier programming lets us quickly build DIY solutions tailored to our own preferences}'' \fix{(follow-up study)}.}
\end{quote}

Importantly, code assistants also expanded participants’ capabilities, \textbf{bridging accessibility gaps} that would otherwise hinder their work. 
They successfully used AI to recognize pin diagrams of electronic components and convert code screenshots from blogs into editable code---tasks that would be extremely difficult without AI support.
One prominent domain was \textbf{UI development}, which prior work has described as particularly challenging due to its inherently visual nature~\cite{pandey2021understanding,pandey2022accessibility}. 
\fix{For instance, P10 stated,}
\begin{quote}
    ``\emph{I primarily worked on backend development in the past, as my visual impairment made it difficult for me to handle UI tasks effectively. [...] I turned user feedback into prompts for Copilot to make changes, then asked it to check its generated code and sent interface screenshots for further inspection. I also double-checked the code myself. Once I was confident, I asked a sighted colleague to verify the interface and made further adjustments based on their suggestions. [...] It can generate a usable interface, even if the visual design may be basic. Still, this has made my workflow much easier. Before, when I wrote UIs, I focused solely on preventing element overlap---a task that required significant mental energy to determine proper component placement. Now, with Copilot, I no longer have to handle this challenging process alone}'' \fix{(follow-up study)}.
\end{quote}

\subsection{\fix{Challenges in Complex Human-AI Interaction Workflows}}
\label{sec-4-challenges}
\begin{figure}[htbp]
    \centering
    \includegraphics[width=\linewidth]{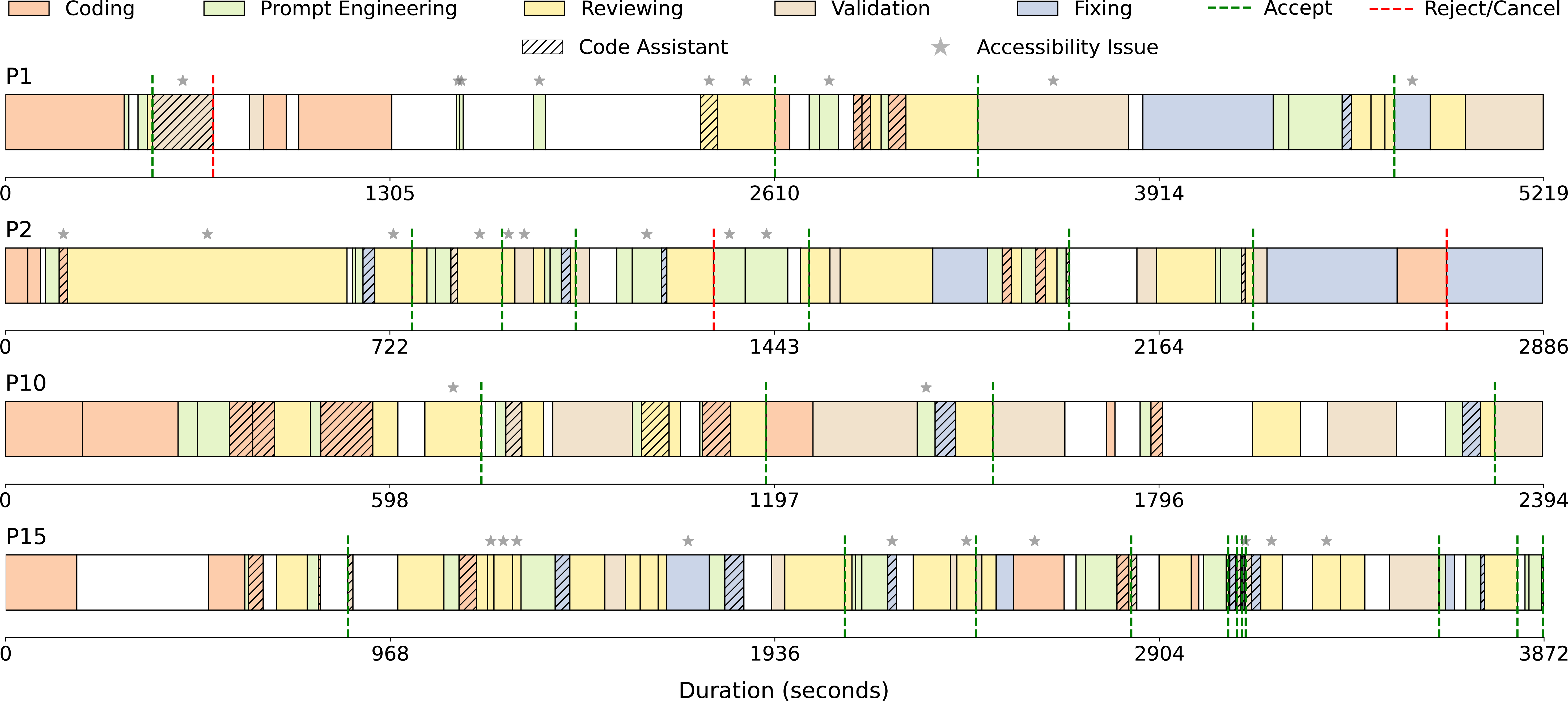}
    \caption{Four users' timelines---P1 (task: T2), P2 (task: T4), P10 (task: T1), and P15 (task: T3)---during their programming task. Code Assistant indicates when the assistant generated a response or performed an action after receiving a user's prompt or acceptance. Accessibility Issues denote points where users encountered difficulties that hindered smooth operation during the activity. Blank intervals represent periods of non-programming activities (e.g., seeking assistance for issues, troubleshooting network issues).}
    \label{fig:timeline}
    \Description{Timelines show heterogeneous, non-linear workflows with frequent reviewing and accessibility interruptions. Each track alternates activity segments (coding, prompt engineering, reviewing, validation, fixing) with assistant action markers. Activity ordering differs across participants, and human oversight clusters tightly around each assistant-generated change, with accessibility issues arising during these activities.}
\end{figure}

Despite the tool's advanced intelligence, the collaboration process remains complex, varies across individuals, and is further shaped by accessibility issues.
\autoref{fig:timeline} illustrates how four representative participants collaborated with the code assistant and encountered challenges, \fix{with complete timelines for all participants provided} in Appendix~\autoref{fig:all_timelines}.
\fix{Analysis of these activities shows that, across all tasks, \textbf{Reviewing} was the most time-consuming, with users spending an average of 11.73 minutes examining the code assistant's responses, followed by \textbf{Prompt Engineering}, averaging 7.51 minutes per user with 8.38 prompts sent.}
\fix{Although users generally find GitHub Copilot accessible, Likert-scale ratings indicate that interaction efficiency still has room for improvement (\autoref{fig:likert_comparison}, Q6-Q7).}
\fix{Building on these observations, in this section we present opportunities and challenges across four areas:} communicating with AI, reviewing AI-generated responses, switching between views, and maintaining awareness of system status and next steps.


\subsubsection{Communicating with AI Code Assistant}
\label{sec-4-communicate}
Participants typically began by configuring the session through the selection of mode, model, and context. After setup, they composed prompts in natural language and submitted them to the assistant. While this process appears straightforward, participants encountered challenges both in basic interaction and in higher-level decisions, such as prompt formulation and model selection.

Copilot's rich set of keyboard shortcuts significantly improved participants' efficiency and sense of control in their workflow. 
However, many participants expressed frustration when \textbf{shortcuts conflicted with their established keyboard habits}, leading to unexpected behaviors or workflow disruptions. 
Nearly all participants struggled with the up and down arrow keys in the input box---rather than moving between lines in a multi-line prompt, these keys cycled through previous commands, often without users realizing it. As P13 put it, ``\emph{It makes editing multi-line prompts cumbersome.}'' 
%
In addition, participants struggled to \textbf{obtain timely feedback about current session settings}, such as selected mode, model, or context. 
As P9 explained, ``\emph{I selected several lines in the editor, but only the file name was announced. I couldn't tell what was actually being used as context.}'' 

Beyond interaction, participants also \textbf{faced challenges in translating intent into effective prompts, particularly \fix{about level of detail}}. When prompts were too concise, Copilot sometimes made incorrect assumptions. For example, P4's brief prompt like ``\emph{please check if there are errors in the code}'' \fix{(Claude 3.7)} led Copilot to suggest changes to calculation rules that were already documented as unchangeable. 
In response, some participants (e.g., P7) provided highly detailed prompts and revised them frequently to avoid misunderstandings, resulting in increased time.

This behavior stemmed from \textbf{uncertainty about the code assistant's capabilities}. As P14 put it, ``\emph{Although there are many models to choose from, I don't know which models are reliable for which tasks, nor when they might make mistakes.}'' 
P10 expressed a desire for models with clearly defined capabilities:
\begin{quote}
    ``\emph{It would be better if there were a dedicated model for writing documentation. Now it's hard to decide which model is best. If I could clearly know which model performs best for which task, I would be more confident in my choice}'' \fix{(follow-up study)}.
\end{quote}

Over time, participants \textbf{developed their own strategies} for crafting prompts to communicate more effectively.
P8 shared, ``\emph{I now list requirements one by one, and specify the expected outcome. I try to communicate all my needs as clearly as possible. The cycle of checking responses and revising prompts is time-consuming}'' \fix{(follow-up study)}.
\fix{Experienced AI code assistant user} P10 emphasized the importance of actively guiding the assistant, 
\begin{quote}
    ``\emph{It should be \textbf{you use AI, not AI uses you}. At first, I would just let the model do its thing. But over time, I realized that doesn't work, the model can easily go off track if you don't guide it. Now, when I have a complex task, I first ask it to clarify and confirm my needs, and only then do I let it modify the code or I make the changes myself. I find I need to give explicit instructions to keep things under control. I'll say things like `do not modify the code yet' or `make changes step by step' to make sure it doesn't rush ahead and change things based on its own assumptions instead of mine.}''
\end{quote}

\subsubsection{Reviewing Generated Responses}
\label{sec-4-review}
Reviewing the outputs from code assistants---both code snippets and textual explanations---is critical for users to understand and evaluate the assistant's suggestions. 
However, participants reported several challenges in this process. 



For programmers, reviewing code directly in the code editor (\autoref{fig:copilot} (1)) is intuitive. 
After code edits are applied, users typically return to the editor to review file modifications.
\textbf{Meaningful comments in generated code} help users understand the logic and reduce the need to switch between views. 
%
%
However, participants expressed a need to \textbf{clearly understand \fix{the specific changes made to the code}}.
P3 explained,
\begin{quote}
    ``\emph{When browsing code differences in the diff view, the mixing of unchanged, deleted, and added code makes it hard to follow the changes. Since I can't scan, I have to remember what I just heard. If I could switch between old and new code with a shortcut, it would help me understand the changes better.}''
\end{quote}
Another challenge was \textbf{tracking the number, location, and status of changed code blocks}. 
P11 said, ``\emph{If there are many code block changes, I get confused and don't know how many I have not reviewed.}'' 
P3 also mentioned, 
\begin{quote}
    ``\emph{Automatically jumping to the next change after I accept a code change confuses me and makes me lose track of where I am. I wish I could control whether to jump, and if it would automatically read the code line after jumping, that would be more convenient.}''
\end{quote}

\fix{The message list presents user queries and assistant responses as discrete blocks (\autoref{fig:copilot} (2)).}
%
\fix{Participants appreciated the \textbf{structured layout of the message list}, especially as it allowed experienced users to switch seamlessly between \emph{browse mode}\footnote{Browse mode: A screen reader mode that enables users to navigate content using virtual cursors and structural shortcuts (e.g., headings, buttons) without directly interacting with the application.} and \emph{focus mode}\footnote{Focus mode: A screen reader mode where keystrokes are passed directly to the application, allowing users to interact with interface elements and input text in real time.}.}
P1 shared that his preference shifted from the accessible view to the message list, stating, 
\begin{quote}
    ``\emph{The message list has stronger presentation capabilities. I prefer switching the screen reader to browse mode, where I can use familiar shortcuts to jump between headings with \texttt{H}, buttons with \texttt{B}, and code blocks with \texttt{E}, and use the arrow keys to read text}'' \fix{(follow-up study)}.
\end{quote}
\fix{However, inexperienced users and novices found the message list \textbf{difficult to navigate}.}
P9 noted, ``\emph{Replies are sometimes too long, making it inconvenient to browse with arrow keys. If there were a table of contents, it would be much easier.}'' 
P5 said, ``\emph{After opening a complex application like VS Code, I don't switch the screen reader mode anymore. When I navigate to a block in the message list, the information is linear and hard to understand}'' \fix{(follow-up study)}.
%
%
In addition, users pointed out that when using \emph{Agent mode} for multi-turn automatic iteration, all results and changes were presented in a single response block, which was \textbf{not well structured} and made it difficult to locate information. 

Besides, participants had \textbf{higher expectations for the content of responses}.
As P12 described, 
\begin{quote}
    ``\emph{If it could verify whether a referenced function from a non-Python standard library actually exists and accurately tell me which class and function it's in, I'd trust its suggestions more. Now, I suspect it's making things up, which increases my verification time}'' \fix{(GPT-4o)}.
\end{quote}
Other users found that, especially with complex tasks, the model sometimes only partially completed the requirements without making this explicit. For example, P6 submitted an entire report file and expected Copilot to complete all tasks, but only after being notified did the user realize that only the first subtask was completed. P6 said, ``\emph{If it hasn't read everything, it should tell me}'' \fix{(GPT-4o)}.

\subsubsection{Switching Between Views}
\label{sec-4-view}
Interacting with code assistants requires users to switch between many views (e.g., chat panel, editor). Consistent with Flores et al.~\cite{flores2025impact}, \fix{users struggled with view switching, a challenge we found was particularly pronounced in the highly automated \emph{Agent mode}.}

Participants reported that \textbf{managing focus across multiple views} was challenging and sometimes led to confusion or errors. 
During the programming task, P7 accidentally triggered a shortcut that moved the focus from the chat input box to the editor, resulting in text entered into an open file instead. Similarly, P4 described frequent focus jumps between the chat window and the editor while navigating in the inline chat, stating, ``\emph{How can I get back to the previous window?}''

This complexity was evident when participants needed to \textbf{gather information from multiple views} to understand the assistant's feedback.
The introduction of \emph{Agent mode}, which allows automatic code modifications and execution, further exacerbated these challenges. While \emph{Agent mode} uses the built-in terminal, users had to \textbf{manage an additional but unfamiliar view}. 
\fix{However,} most participants \fix{rarely used} the built-in terminal, as they were concerned about focus traps or found switching between views cumbersome.
For instance, after automatic execution \fix{(GPT-4o)}, P15 \fix{switched} between the editor and the message list to review code changes and explanations\fix{, and only noticed the failure in the terminal after the interviewer pointed it out}. 

Additionally, participants found it difficult to \textbf{verify results across multiple views}. 
For example, after printing data analysis results in the terminal, P3 and P8 tried to use the terminal output as context, hoping Copilot would insert the results into reports.
However, the assistant inserted hallucinated data instead of the actual output. After being reminded, users spent significant time understanding the error and frequently switched between different views to compare the data. 
P8 remarked, ``\emph{How can we verify it? It can lie}'' \fix{(Claude 3.7)}.
P3 added, ``\emph{If it could write the correct result directly into the file, it would save me a lot of steps. But if I still have to check like this, I'd rather copy and paste it myself}'' \fix{(GPT-4o)}.

\subsubsection{System Status and Next Steps}
\label{sec-4-status}

After submitting a request, users must wait for the assistant's response. \textbf{Timely status notifications} are essential for helping users understand the generation state and reduce uncertainty while waiting. As P10 described, 
\begin{quote}
    ``\emph{Cursor does not provide status notifications well. I keep checking to see if the response is generated. Copilot, on the other hand, plays a continuous sound while generating a response, so I can work on other tasks in parallel and return when I hear it is done.}''
\end{quote}

However, \textbf{status notifications are not clear enough}, \fix{often leaving them unsure of what action to take next}.
For example, when \emph{Agent mode} generates terminal commands and waits for user confirmation before executing them, the system announces ``\emph{Action required: Run command in terminal,}'' but then continues to say ``\emph{Progress,}'' making it unclear that the assistant is waiting for the user to take action.
After hearing such notifications, P4 asked, ``\emph{Do I need to execute the command now? What exactly is the command?}'' 
P7 noted, ``\emph{Status notifications are brief. If I miss them, I don't know what happened.}'' 
To keep the study moving forward, the interviewer would relay status information to the participant whenever they asked. In response, P3 remarked, ``\emph{You are my accessibility feature.}''

Beyond clearer instructions, participants also \textbf{wanted greater transparency in the process}, \fix{so they could understand what the assistant was doing at each step and} judge earlier whether the assistant's actions aligned with their expectations. 
During response generation, the system continuously announces ``\emph{Working}'' and ``\emph{Progress}'' accompanied by sound cues to indicate ongoing processing. 
P8 commented, ``\emph{Visually, I can vaguely see text being generated, but there is no timely notification. Only after the generation is complete do I hear details.}'' 
P4 also felt, ``\emph{The notifications during execution are not specific enough. Whether it is executing code or starting to make fixes?}''

\subsection{\fix{Accessibility Features: Trade-offs and User Perceptions}}
\label{sec-4-accessibility}
\fix{Participants generally appreciated the accessibility features in GitHub Copilot for their convenience, but also highlighted limitations, revealing the subtle trade-offs in accessible interface design.}

\subsubsection{\fix{Accessible View: Ease of Use vs. Information Completeness}}

The accessible view (\autoref{fig:copilot} (3)) in Copilot is designed to alleviate navigation difficulties in the message list. It presents information in a single, plain-text window, allowing users to browse all details using only basic arrow key navigation. 
Participants praised these accessibility features for their \textbf{convenience and ease of use}.
For example, P1 shared, 
\begin{quote}
    ``\emph{The accessible view is great because I don't have to switch between different areas. I just open it right from where I'm typing, read the response, close it, and I'm back to typing again. Moreover, When I open the accessible view, I can be sure my focus is at the top.}'' 
\end{quote}
P3 also stated, ``\emph{Compared to the message list, I use the accessible view more often. Navigating the message list requires too many key operations}'' \fix{(follow-up study)}.

However, \fix{this convenience comes with trade-offs.}
The accessible view \fix{sometimes} \textbf{\fix{reduces} information expressiveness and completeness}.
\fix{In the follow-up study, P1's preference shifted from the accessible view to the message list, as the latter offered greater expressiveness and ease of navigation.}
P6 noted that ``\emph{the accessible view sometimes includes markup language or presents formatted information like tables in ways that are difficult to interpret.}''
Additionally, P5 found that ``\emph{the accessible view only showed the changed code, without indicating which files were affected.}''
Both P2 and P7 suggested that automatic updates would improve usability, as the accessible view cannot show newly generated content without manual refreshing.

\subsubsection{\fix{Automatic Playback and Status Notifications: Convenience vs. Disruption}}
\label{sec-4-autoplayback}
\fix{While automatic playback can facilitate reviewing content, it can also \textbf{disrupt users' concetrations}.}
P1 said, ``\emph{Just a sound cue is enough. Automatic playback disrupts my thinking. Also, linear reading makes complex responses difficult to understand}'' \fix{(follow-up study)}.
%
\fix{Similarly,} some participants found continuous status notifications \textbf{distracting}. 
P14 said, ``\emph{When I switch to browse mode in the message list to review, the `progress' notification interrupts what I'm listening to, so I have to go back and listen to the same sentence again}'' \fix{(follow-up study)}.
P5 shared, ``\emph{I work on other tasks in parallel. In such cases, notifications break my attention, and I try to ignore sound cues. I wish it would notify me only when my action is required}'' \fix{(follow-up study)}.

\subsubsection{\fix{Sound Cues for Code Changes: Helpfulness vs. Cognitive Load}}
Participants \textbf{valued well-designed sound cues} for code modifications~\cite{seo2023coding}.
P13 commented, ``\emph{Having more types of cues, like sound cues, really helps me understand changes better. If there are too many textual cues mixed in, it might actually interfere with my understanding.}''

However, some participants found them \textbf{confusing or difficult to distinguish}, which occasionally hindered their ability to interpret code changes accurately. 
P7 shared, ``\emph{It's easy to mix up the sound cues and forget what they mean}.''
P5 added, ``\emph{I wish there were more direct explanations}'' \fix{(follow-up study)}.
P13, who praised sound cues, also mentioned, 
\begin{quote}
    ``\emph{At first, I couldn't tell the different sound cues apart, but since I'm familiar with the code, I could roughly guess whether it was an addition, deletion, or modification. Over time, I gradually got used to them and came to appreciate the benefits of sound cues.}''
\end{quote}

\subsection{Feature Preferences: Automation vs. Control}
\label{sec-4-automation}
\begin{figure}[htbp]
    \centering
    \includegraphics[width=0.9\linewidth]{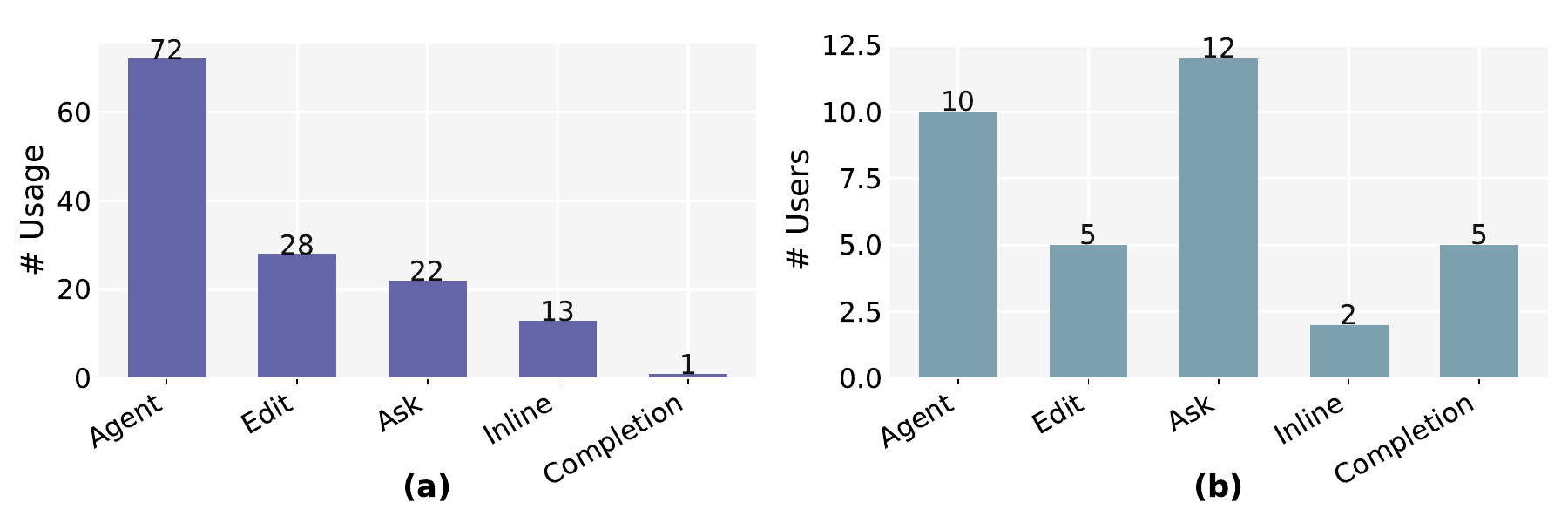}
    \vspace{-1em}
    \caption{Usage and user preference for different Copilot features. (a) Initial study: Number of times each feature was used during programming tasks. (b) Follow-up study: Number of participants reporting frequent use of each feature.}
    \label{fig:fig_mode}
    \Description{Feature usage shifts from automation-first to control-oriented across phases. Panel (a) bar chart shows that during the programming task, usage counts rank Agent mode highest, then Edit mode, then Ask mode. Panel (b) bar chart shows that follow-up self-reports invert the pattern. Ask mode is now most commonly used, Agent mode drops to second, Edit mode third.}
\end{figure}

\fix{After two weeks of exploration, participants generally became more familiar with Copilot's interactions and developed preferences for different features. Observing their experiences across different modes, we identified a trade-off between automation and user control (see \autoref{fig:likert_comparison}, Q8-Q9 for their Likert scores).}
As shown in~\autoref{fig:fig_mode}(a), the \textbf{most advanced and automated} \emph{Agent mode} was used most frequently.
Many participants were eager to try the highly automated \emph{Agent mode}. As P1 commented, ``\emph{I find Agent mode to be the most powerful, as it includes all the features of Edit mode. I feel that just using this mode is sufficient for my needs.}''
P10 described \emph{Agent mode} as ``\emph{powerful}'' but noted its tendency to ``\emph{randomly change}'' content that was not intended to be altered, reducing confidence in AI oversight. 
However, in the follow-up study (\autoref{fig:fig_mode}(b)), usage patterns shifted, with participants increasingly favoring the \textbf{safer} \emph{Ask mode}. 
This shift reflected participants' growing desire for control.

Users need to retain control in automated AI systems.
\fix{In the follow-up study,} P2, P10, and P12 \fix{all reported preferring} \emph{Edit mode} but rarely used \emph{Agent mode}, \fix{expressing concerns about \textbf{losing control}.}
\fix{P10 articulated this loss of control more concretely---spanning the \textit{scope of context}, the \textit{task flow}, and even \textit{resource consumption}---stating,}
\begin{quote}
    ``\emph{Agent mode is slow and searches files in ways misaligned with my intent. It performs unnecessary steps that consume my usage quota. Because I already know exactly which files I need, Edit mode is sufficient for my workflow}'' \fix{(follow-up study)}.
\end{quote}
For others, even \emph{Edit mode} felt risky. P6 and P7 rarely used either \emph{Edit mode} or \emph{Agent mode}, expressing a strong sense of \textbf{discomfort} with the assistant making direct changes to their files. 
\fix{This discomfort was further exacerbated by challenges when interacting with more automated modes (see \autoref{sec-4-view} and \autoref{sec-4-status}).} As P6 described, 
\begin{quote}
    ``\emph{Agent mode is hard to manage. It edits my files, but it's not easy for me to review. I'm never sure if I've checked everything, especially when multiple blocks are modified. The status notifications also confuse me, and I don't know what step it's on}'' \fix{(follow-up study)}. 
\end{quote}
P13, despite frequently using \emph{Agent mode}, \fix{also expressed concerns about its lack of controllability due to unfamiliarity with the built-in terminal, noting}, 
\begin{quote}
    ``\emph{I only use Agent mode to generate code, not to execute commands automatically. Letting it execute feels \textbf{unreliable} and doesn't fit my habit of using external terminal tools; running code myself feels more familiar and gives me more confidence}'' \fix{(follow-up study)}. 
\end{quote}

Most participants viewed \emph{Ask mode} as a \textbf{safer} choice because it required explicit confirmation before modifying files. As P11 explained, ``\emph{I like to discuss my plan or ask for explanations in Ask mode first, before letting Copilot touch my files}'' \fix{(follow-up study)}. This approach allowed them to use \emph{Ask mode} for brainstorming, learning, or clarifying requirements before switching to Edit or Agent mode for implementing actual code changes.
\fix{Similarly,} inline chat earned praise from P1 and P6 for offering \textbf{precise control}. P6 noted, 
\begin{quote}
    ``\emph{With inline chat, I can select specific code and make small changes. The diff view opens automatically, making it easy to review code differences and giving me a greater sense of control}'' \fix{(follow-up study)}.
\end{quote}
P1 agreed \fix{but highlighted a desire for more control, explaining}: ``\emph{I felt that inline chat was \textbf{less stable} because of how the window floats. A fixed window feels more stable}'' \fix{(follow-up study)}. 

Code completion was less popular among our participants.
While some initially felt overwhelmed or distracted~\cite{flores2025impact}, many eventually adapted and found the auditory cues less disruptive \fix{(see \autoref{fig:likert_comparison}, Q10)}. Others encountered practical issues, such as the \texttt{Tab} key conflicting with indentation, uncertainty about when suggestions would appear, and less manual coding resulting in fewer code completions.

\subsection{Barriers to Learning for Screen Reader Users}
\label{sec-4-learning}
\fix{Most} screen reader programmers \fix{in our study had} limited exposure to \fix{code assistants}.
However, once introduced, participants expressed strong interest and willingness to use code assistants, recognizing their benefits across development tasks (\autoref{sec-4-cases}), consistent with prior findings~\cite{flores2025impact}.
This gap between interest and adoption highlights several critical barriers to learning and onboarding for screen reader users.
\fix{Based on participants' learning processes in the initial study and their feedback during the two-phase interviews, we identified three main challenges they encountered.}

A key barrier to learning is \textbf{the inaccessibility of code assistant tools}, a long-standing issue in the accessibility of programming tools~\cite{mountapmbeme2022addressing, pandey2021understanding}.
Participants who had used Cursor and Cline reported significant challenges with these tools. P14 described, 
\begin{quote}
    ``\emph{Some UI elements on the Cursor, like code blocks and chat input fields, lack clear labels or distinction in their types. For example, both appear as `Input' components, making it difficult for me to tell them apart. To address this, I write a custom NVDA add-on to process these elements and add more meaningful labels. Additionally, when the code assistant makes changes to my files, it's often hard to understand what modifications were made. I rely on git diff to see the changes, which is not always an intuitive solution.}''
\end{quote} 

\textbf{The complexity of elements in interfaces} also increases cognitive load and anxiety when exploring tool functionalities with a screen reader.
P5 noted, 
``\emph{For such complex software, I don't attempt to navigate through all the elements---there are too many. I worry about getting stuck in an element and being unable to exit}'' \fix{(follow-up study)}.
\fix{Participants' difficulties in exploring complex interfaces hindered their ability to discover new features.}
For example, although Copilot includes an undo function to revert code changes 
, many participants were still unaware of its existence after two weeks of exploration and asked whether such a feature was available \fix{in the follow-up study}.
When informed about the presence of an undo button within the message block, P5 responded, ``\emph{That feature is hard to find. I rarely go back to my own messages. If it had been placed in the assistant's response block, I might have noticed it}'' \fix{(follow-up study)}.
Instead, he demonstrated an alternative interaction strategy: ``\emph{I just asked Copilot to undo using natural language, and it seemed to work}'' \fix{(follow-up study)}.
P1 \fix{experienced difficulties navigating and maintaining context when exploring new and complex interfaces, a challenge} compounded by the combined accessibility and compatibility issues of using a screen reader, an IDE, and the operating system.

\textbf{The lack of accessible learning resources} often leads to screen reader users spending considerably more time learning new tools. Key gaps include screen reader-specific documentation---particularly for new features---and community-based technical sharing. Similar challenges have also been observed in the context of productivity applications~\cite{perera2025sky}.
As P14 recalled, ``\emph{When Cursor became popular, none of the recommendations mentioned how to operate it with a screen reader. I had to rely on general tutorials and my own experience to figure things out.}'' P13 further elaborated, 
\begin{quote}
    ``\emph{I usually start by looking for official documentation, but that depends on whether the developer provides it. After that, I browse technical blogs, whether text or video. However, most of these are aimed at sighted users, so I typically need to convert them into an accessible format myself. For example, if it says `click here,' I try to figure out where that is based on the context. If the software design follows familiar patterns, it's easier to understand, but if not, it can be pretty challenging}'' \fix{(follow-up study)}.
\end{quote}



\section{Discussion}
\label{sec-discussion}

\subsection{\fix{Reflections on Vibe Coding}}

\fix{
While vibe coding was originally proposed to describe a programming style in which developers primarily guide AI rather than directly write code, enabled by highly agentic LLM tools~\cite{karpathy2025vibecoding}, its precise boundaries remain fluid and socially negotiated~\cite{silverstein2003indexical,sarkar2025vibecoding,pimenova2025good}. 
Rather than assuming vibe coding as a fixed workflow, we designed our study to investigate how screen reader programmers could engage with this emerging paradigm. To this end, we selected GitHub Copilot with Agent mode, which supports the natural-language, agentic interactions characteristic of vibe coding while fitting into participants’ existing workflows. This setup allowed us to analyze how participants interacted with advanced automation features and to identify factors that facilitated or hindered their engagement in vibe-coding-like practices.
}

\fix{
Our findings show that screen reader programmers clearly recognize the value of advanced, intelligent code assistants (\autoref{sec-4-cases}). 
As identified in prior studies~\cite{Liang2024survey, dakhel2023github,kazemitabaar2023studying, puryear2022github}, code assistants can enhance efficiency and facilitate learning; our participants echoed these observations. 
Beyond these benefits, we also found that multimodal capabilities introduced new accessibility advantages---such as interpreting code embedded in images or assisting with UI development, tasks previously considered highly challenging~\cite{pandey2022accessibility, ferrari2021accessible}.
}

\fix{
However, our observations suggest that participants only partially engaged in the kind of high-level, intent-driven workflow described by Karpathy~\cite{karpathy2025vibecoding}.
In the initial study, we observed that participants reduced manual coding time and shifted substantial effort toward prompt engineering, aligning with descriptions of vibe coding as natural-language, conversational programming~\cite{sarkar2025vibecoding,pimenova2025good}. 
Yet trust in AI-generated changes was insufficient~\cite{adnin2024look,alharbi2024misfitting} for participants to fully ``forget the code'' (\autoref{fig:likert_comparison}, Q4).
As shown in~\autoref{fig:fig_time}, they frequently validated outputs by running programs and tests but still invested significant time reviewing code and suggestions. 
In the follow-up study, we further observed a strong desire to maintain control and accountability (\autoref{sec-4-automation})---concerns that align with prior findings on code assistants, particularly in professional settings~\cite{examining2025,flores2025impact}.
}

\fix{
In addition, we found that participants' interactions with the code assistant were not always smooth (\autoref{sec-4-challenges}). 
Difficulties in articulating intent and reviewing responses---consistent with prior work~\cite{bansal2024challenges,adnin2024look,flores2025impact,alharbi2024misfitting}---were amplified for screen reader programmers in more complex coding scenarios.
Due to the lack of real-time visual feedback from the assistant, participants experienced increased cognitive and navigational burdens and exacerbated uncertainty about changes, system state, and next steps. 
Agent mode, which automates more operations and makes broader modifications in the workspace, introduced additional friction related to perceiving AI status, understanding its processes, and interpreting outputs, making it more difficult for participants to maintain a sense of control. 
Moreover, as AI tools evolve rapidly, we observed challenges in maintaining up-to-date skills (\autoref{sec-4-learning}), highlighting an ongoing accessibility issue that could widen existing skill and opportunity gaps~\cite{perera2025sky,storer2021s}.
}

\fix{
Overall, our findings suggest that while advanced agentic tools can support aspects of vibe-coding-like workflows, substantial accessibility, trust, and coordination barriers prevent screen reader programmers from fully participating in this paradigm. These insights highlight the need for more inclusive design of agentic code assistants to ensure that screen reader programmers can fully benefit from evolving programming practices.
}

\subsection{Toward a Sense of Control}
Our findings indicate that while automation enhances efficiency and reduces effort, preserving control over the interface and the AI collaboration process is equally essential.
Users emphasized the need for transparency and predictability---both in anticipating the consequences of their own actions and in understanding how the AI will behave. \fix{Consistent with prior work showing that screen reader programmers often experience frustration and a loss of control in complex IDEs~\cite{albusays2017interviews,albusays2016eliciting}, our results reveal that the dynamically changing environment introduced by AI assistance further amplifies these challenges.
Participants not only struggled with uncertainty about their current focus but also with ambiguity regarding what the AI had done or was in the process of doing. 
Supporting users in maintaining a sense of control has long been a central concern in HCI and accessibility research~\cite{coyle2012did,limerick2014experience,legaspi2024sense,ranganeni2023exploring}. 
Building on our findings, the following sections examine how increased feature complexity and greater automation can introduce new risks for screen reader programmers, highlighting the need for careful consideration.}

\textbf{Customization: Less or More}.
To meet diverse user needs, software tools have introduced increasingly rich features. However, for screen reader users, more features don't always mean better usability.
In~\autoref{sec-4-accessibility}, users showed varying preferences between the structured message list (\emph{more}) and the plain-text accessible view (\emph{less}). 
While the message list enables information expressiveness and facilitates structured navigation, users unfamiliar with shortcuts found it challenging and frustrating.
The simpler accessible view, though limited to line-by-line navigation, offered greater clarity and control. 
This trade-off extends to users' choice of development environments. P4, for example, preferred reading code in Notepad, finding VS Code ``too complex'' with ``too many interaction steps.''
Prior studies~\cite{Mealin2012, albusays2016eliciting} similarly report that many screen reader users favor simple editors over feature-rich IDEs, as complex interfaces often hinder accessibility due to increased UI elements and shortcut conflicts.
%
Some users valued streamlined workflows and task efficiency---calling for \emph{more} functional support.
However, excessive complexity and unpredictability often caused overload---calling for \emph{less} cognitive and interactional burden. 
This tension highlights the need for customizable, user-centered design, as echoed in prior work~\cite{perera2025sky,jones2024customization}. 
Enabling systems to adaptively adjust interaction complexity based on user preferences may be key to balancing usability and efficiency.
%

\textbf{Greater Automation, Increased Risks}.
While users generally appreciated advanced features such as \emph{Agent mode}, these highly automated interactions also introduced new challenges---particularly concerning transparency, control, and unintended changes. As noted in~\autoref{sec-4-status}, participants emphasized the need for clear and timely status updates during task execution to track AI actions. Similarly, in~\autoref{sec-4-review}, users emphasized the need to examine and verify outcomes, especially for direct file modifications.
This reflects a classic automation paradox: increased system autonomy can diminish users' sense of control and raise concerns about potential errors. To cope with these risks, some participants reported adopting compensatory strategies, such as version control with Git (P2), to maintain oversight and recoverability.
These findings echo prior research on human-AI collaboration~\cite{wen2022sense, kocielnik2019will, chen2022trauma, bansal2024challenges}, which stresses the importance of transparency, user agency, and trust.


\subsection{For the Future: Towards Accessible Human-AI Collaboration Design}
Our findings reveal the unique needs \fix{and challenges of screen reader programmers when interacting with AI code assistants.}
Similar concerns also arise in other AI tools with complex UIs and conversational interactions~\cite{chaves2021should, casheekar2024contemporary}, such as Microsoft Copilot~\cite{microsoftcopilot}, Google Gemini in Workspace~\cite{googleworkspacegemini}, and Notion AI~\cite{notionai}.
As these tools become increasingly widespread, addressing human-AI interaction challenges is critical~\cite{bansal2024challenges}.
In light of prior work on generative AI design~\cite{weisz2024design, Tahaei2023Human}, we propose \fix{\textit{high-level principles} for accessible human-AI collaboration, followed by \textit{concrete recommendations} grounded in the observations of our study.}
\subsubsection{\fix{High-Level Principles}}
\begin{itemize}[leftmargin=\parindent]
    \item \textbf{Embrace Interaction Simplicity.}
    Minimize cognitive load through streamlined and predictable interaction patterns.
    Complex interfaces create additional barriers for screen reader users who experience interface content linearly through audio feedback. AI systems should prioritize simple, efficient interaction with consistent behaviors across features, \fix{enabling users to form} accurate mental models and reducing learning overhead.
    
    \item \textbf{Maintain Transparent Communication.}
    Ensure that AI processes and outcomes are accessible beyond visual interfaces.
    Screen reader users require comprehensive, non-visual feedback about both AI actions and results, as current interfaces rely mostly on visual-only  indicators. Transparency encompasses real-time \fix{status updates}, clear explanations of outputs, and accessible methods for users to verify, understand, and act upon AI-generated results.
    
    
    \item \textbf{Design for Inclusive Learning Journeys.}
    \fix{Support users in developing both operational skills and strategic understanding of how to collaborate effectively with AI assistants.}
    Traditional documentation and visual tutorials often fail to address the learning needs of screen reader users. AI systems should support personalized \fix{and accessible learning pathways that help} users build procedural knowledge and strategic understanding of human-AI collaboration.
\end{itemize}

\subsubsection{\fix{Concrete Recommendations}}
\fix{Based on these principles and our findings, we provide specific recommendations to guide the design of accessible AI assistants for screen reader users:}


\textbf{\fix{1. Interaction and Navigation}}
\begin{itemize}[left=12pt]
    \item Design consistent and predictable keyboard shortcuts to efficiently access key functions and reduce unnecessary navigation\fix{, aligning with prior recommendations for standardization in assistants~\cite{perera2025sky}}. \fix{This addresses participants' difficulties with unintended actions that disrupt their workflow (\autoref{sec-4-communicate}) and hinder learning to effectively use the AI tool (\autoref{sec-4-learning}).}
    \item \fix{Structure and integrate related messages and content to support efficient navigation~\cite{borodin2010more,jordan2024information}. For example, group messages with headings or a table of contents and present related code, comments, or commands with results, helping users maintain context and quickly locate information (\autoref{sec-4-review}, \autoref{sec-4-view}).}
    \item Support accessible change tracking\fix{, including textual or audio cues for the} scope (e.g., number of edits), context (e.g., location and status of code blocks), and content (e.g., what was modified). \fix{Participants reported difficulty understanding changes, which hindered their confidence in accepting or refining suggestions (\autoref{sec-4-review}).}
    \item \fix{Provide mechanisms to reduce interruptions (\autoref{sec-4-autoplayback}), such as a ``Do Not Disturb'' mode, and offer on-demand status checks to maintain focus and situational awareness (\autoref{sec-4-status}).}
\end{itemize}

\textbf{\fix{2. Transparency and Verifiability}}
\begin{itemize}[left=12pt]
    \item When prompts are unclear, AI assistants should proactively seek clarification by asking users to specify their intent or provide more additional details\fix{~\cite{tang2025clarifying,zhang2025clarify}}, thus supporting better mutual understanding \fix{(\autoref{sec-4-communicate})}.
    \item Improve the verifiability of \fix{AI responses by} providing references, \fix{explanations, or evidence, allowing} users to \fix{assess the reliability of suggestions. Participants struggled to review and trust outputs without supporting information (\autoref{sec-4-review}), echoing findings from prior work~\cite{adnin2024look,perera2025sky}.}
    \item \fix{Make AI processes more transparent by providing clear and accessible status notifications that inform users what} the AI is doing and which actions require their input. \fix{Our findings indicate that participants relied on timely and understandable status feedback to maintain trust, verify AI behavior, and make informed decisions (\autoref{sec-4-status}).}
\end{itemize}

\textbf{\fix{3. Learning and Customization}}
\begin{itemize}[left=12pt]
    
    \item Provide guidance for model selection or automatically recommend suitable models based on user intent\fix{~\cite{luo2020metaselector}, supporting participants who experienced confusion in choosing appropriate AI models (\autoref{sec-4-communicate})}.  
    \item \fix{Offer optional accessibility features, which allow users to personalize the interface according to their needs (\autoref{sec-4-accessibility}) and are often emphasized in existing research on accessible development tools~\cite{potluri2018codetalk,potluri2022codewalk}.}  
    \item Leverage AI's conversational capabilities to guide users through personalized learning journeys, helping them acquire both operational and conceptual knowledge. \fix{Participants reported that existing learning resources are limited and often inaccessible (\autoref{sec-4-learning}), and expressed a desire for guidance similar to what researchers provided during GitHub Copilot onboarding, to help them navigate new features or difficulties.}
\end{itemize}

\subsection{Limitation and Future Work}


\fix{
Our study involved 16 participants, although this number is above the median reported in accessibility studies for BLV individuals~\cite{mack2021we}, it can be increased in the future for broader generalizability.
The study focused mainly on screen reader programmers in China, so behaviors may differ from those of screen reader users in other regions or cultural contexts, as cultural norms, programming practices, and accessibility standards can shape user experiences.
While the study included two female participants, the overall gender distribution remained skewed toward male participants, }reflecting broader challenges in achieving gender diversity within HCI and accessibility research, particularly in programming~\cite{pandey2021understanding, datavoidant2022}.
Future work should explore more diverse and inclusive user backgrounds and usage contexts to provide a more generalized paradigm for accessible human-AI interaction. 

\fix{To minimize participant burden and protect their privacy, our longitudinal study relied on self-reported diary entries and follow-up interviews to capture participants' experiences over time. While this approach effectively documented participants' evolving subjective perceptions, it may not fully capture all interaction details with Copilot. Future work could adopt more detailed or systematic data collection methods, while incorporating measures to anonymize or remove personally identifiable information~\cite{abbott2019local}, to better examine users' long-term usage patterns, behaviors, and strategies when interacting with AI code assistants.
}

\fix{Our study highlighted several patterns in screen reader programmers' interactions with AI code assistants; however, the effects of individual differences and model variability were beyond the scope of the current study. These factors may still shape user experiences and merit further investigation. Future work could investigate how prior AI experience, programming proficiency, or screen reader usage shape interactions, and how different models influence behavior and perceived utility, providing guidance for designing more effective and inclusive code assistants.}

\fix{Finally, while vibe coding is lowering the barrier to entry and demonstrating the possibility for more everyday screen reader users to engage in programming, enabling do-it-yourself solutions to their daily life and work needs~\cite{cha2025dilemma,herskovitz2023hacking}, our study did not fully explore how these users experience advanced code assistants. Future work should investigate how screen reader users interact with increasingly capable code assistants, whether they can effectively use them to address personal or practical needs, and what factors influence their adoption and sustained use.}




\section{Conclusion}
\label{sec-conclusion}

This paper presents a two-week, three-phase longitudinal study exploring screen reader users' engagement with advanced AI code assistants and their impact on programming practice. 
Our research uncovers how AI code assistants enhance programming capabilities and bridge accessibility gaps.
While AI code assistants offer clear benefits, users still need better support for communication, output review, view management, situational awareness, maintaining control, and learning advanced features. Drawing on these findings, we propose \fix{design principles and concrete recommendations} to guide future development of AI-assisted tools that empower screen reader users and foster accessible, inclusive human-AI collaboration.
We anticipate that our work can inspire advanced research and engineering efforts to promote an inclusive era of future AI-assisted tools, enabling all users to benefit from the ``\emph{vibe}'' brought by AI.


\bibliographystyle{ACM-Reference-Format}
\bibliography{main}

\appendix
\clearpage
\setcounter{figure}{0} 
\renewcommand\thefigure{\Alph{figure}} 

\setcounter{table}{0}
\renewcommand\thetable{\Alph{table}}

\section*{\fix{Appendix 1: List of Programming Tasks during Initial study}}
\fix{We present four programming tasks used in the initial study in~\autoref{tab:tasks}.}
\begin{table}[ht]
  \centering
  \caption{Overview of four programming tasks simulating common programming scenarios: requirements analysis, code comprehension, testing, debugging, and documentation.}
  \label{tab:tasks}
  \resizebox{\textwidth}{!}{
    \begin{tabular}{@{}p{0.03\textwidth} p{0.17\textwidth} p{0.14\textwidth} p{0.62\textwidth}@{}} 
      \toprule \textbf{ID} & \textbf{Task}  & \textbf{Files} & \textbf{Description} \\
      \midrule 
      T1 & Chat Server~\cite{barke2023grounded} & 3 source files \newline + 1 README & \textbf{1.} Implement server-side logic of a chat application, involving a small state machine. \newline \textbf{2.} Add a \emph{quit} command involving changes to two files.  \\
    \hline
      \rule{0pt}{3ex}T2 & Chat Client~\cite{barke2023grounded} & 3 source files \newline + 1 README & \textbf{1.} Implement networking code for a chat application, using a custom cryptographic API and standard but often unfamiliar socket API. \newline \textbf{2.} Same as the second point of \emph{T1}.   \\
\hline
      \rule{0pt}{3ex}T3 & Data Analysis~\cite{zhang2024benchmarking} & 1 csv file \newline + 1 report.md 
      & \textbf{1.} Directly manipulate the data in the CSV file and populate the results into the report according to the specified table format. \newline
        \textbf{2.} Perform data cleaning, conduct correlation analysis, and insert the processed values back into the report. \\
\hline
      \rule{0pt}{3ex}T4 & Calculator~\cite{mozannar2024realhumaneval} & 1 source file & Identify and fix issues in a calculator class, including missing edge case handling and logical errors such as failing to handle invalid operations like division by zero, and incorrect undo logic. The original Python version fails \textbf{10} test cases, while the JavaScript version fails \textbf{9}, as a Python-specific conversion error does not occur in JavaScript syntax. \\
      \bottomrule
        \end{tabular}
  }
  \Description{Four programming tasks covering server logic, client networking with crypto and sockets, data cleaning and analysis, and debugging a faulty calculator.
  Each task specifies involved files and two core subtasks (except T4 which enumerates bug categories).
  They collectively span requirements analysis, comprehension, implementation, data processing, and error fixing.}
\end{table}

\section*{Appendix 2: List of Questions during Interviews}
\label{sec: app-question}

We present the full phrase of all our questions used in the interview.

\autoref{tab:initial} shows the semi-structured questions asked in the initial interview. 

\autoref{tab:follow} shows the semi-structured questions asked in the follow-up interview.

\autoref{tab:likert} shows the questions rated on a 1-7 Likert scale in both the initial and follow-up interviews, where 1 indicates strong disagreement and 7 indicates strong agreement. 

\begin{table}[h]
\centering
\caption{Likert scale questions' IDs\fix{, labels} and corresponding phrases.}
\label{tab:likert}
\Description{Ten Likert items (Q1-Q10) covering efficiency, learning, predictability, trust, resists overreliance, interaction efficiency, easily reviewed, user control, automation preference and completion distraction.}
\resizebox{\textwidth}{!}{
\begin{tabular}{cp{0.2\textwidth}p{0.7\textwidth}}
\toprule
\textbf{QID} & \fix{\textbf{Label}} & \textbf{Phrase} \\
\midrule
Q1 & \fix{Efficiency} & Copilot's code suggestions make me more efficient in completing programming tasks. \\
\hline
Q2 & \fix{Learning} & Copilot helps me improve my programming skills and inspires new ideas. \\
\hline
Q3 & \fix{Predictability} & I can predict when Copilot will generate useful results. \\
\hline
Q4 & \fix{Trust} & I trust Copilot's code suggestions and am willing to use them directly. \\
\hline
Q5 & \fix{Resists Overreliance} & Copilot's suggestions won't lead me to develop bad coding habits, such as accepting code without fully understanding it. \\
\hline
Q6 & \fix{Interaction Efficiency} & I can efficiently interact with Copilot via keyboard shortcuts or other methods. \\
\hline
Q7 & \fix{Easily Reviewed} & Copilot's suggestions are easy to understand and review using a screen reader. \\
\hline
Q8 & \fix{User Control} & I feel that I am actively leading the collaboration with Copilot, instead of passively following its suggestions. \\
\hline
Q9 & \fix{Automation Preference} & I prefer Copilot to provide suggestions automatically rather than manually triggering them. \\
\hline
Q10 & \fix{Completion Distraction} & Code completion distracts me and makes me feel troubled while programming. \\
\bottomrule
\end{tabular}
}
\end{table}

\begin{table}[ht]
\centering
\caption{Question phrases by category for the initial interview.}
\label{tab:initial}
\Description{Initial interview question set grouped by Background, Thinking process, Interaction practices (learning curve, understanding, evaluation, trust, error handling, feature preference), Copilot's Impact, and Difficulties & Expectations (future adoption, desired improvements).}
\resizebox{\textwidth}{!}{
\begin{tabular}{p{0.15\textwidth}p{0.8\textwidth}}
\toprule
\textbf{Category} & \textbf{Question Phrase} \\
\midrule
\multirow{5}{*}{\makecell[l]{Background \\ (Pre-task)}} & $\cdot$ How long have you been learning programming? \\
 & $\cdot$ Can you share your work experience?  \\
 & $\cdot$ Have you used AI coding assistants before (e.g., GitHub Copilot, ChatGPT)? Which ones? \\
 & $\cdot$ How frequently do you use them? \\
 & $\cdot$ What kind of tasks do you typically use them for? What benefits do you think they provide? \\
\hline
Thinking & $\cdot$ What was your overall thought process when completing the task? \\
\hline
\multirow{11}{*}{Interaction} & $\cdot$ Did you find learning and mastering GitHub Copilot easy or difficult? What aspects influenced this experience? \\
 & $\cdot$ Can you describe how easy or difficult it is to understand Copilot's suggestions? \\
 & $\cdot$ How do you decide which suggestions to accept or reject? How do you evaluate Copilot's suggestions? \\
 & $\cdot$ How much do you trust Copilot? What could Copilot do to make you trust it more? \\
 & $\cdot$ Have you encountered situations where Copilot generated incorrect answers? How did you handle such situations? \\
 & $\cdot$ Among Copilot's different features (code completion, inline chat, ask, edit, agent mode), which do you prefer and why? Does it vary by situation? Any specific examples? Why didn't you use certain modes? \\
\hline
\multirow{5}{*}{Copilot's Impact} & $\cdot$ In your daily programming activities, which tasks or stages do you find most challenging? Please provide examples. \\
 & $\cdot$ Does GitHub Copilot help alleviate the programming challenges you mentioned? Please discuss with specific examples. \\
 & $\cdot$ Additionally, for which tasks do you think Copilot provides the most help? \\
\hline
\multirow{5}{*}{\makecell[l]{Difficulties \\ \& Expectations }} & $\cdot$ What difficulties or challenges have you encountered while using Copilot? If any, what were they? \\
 & $\cdot$ Would you consider using GitHub Copilot long-term? \\
 & $\cdot$ What additional features or improvements would you like Copilot to provide to help you more? \\
\bottomrule
\end{tabular}
}
\end{table}

\begin{table}[h]
\centering
\caption{Question phrases by category for the follow-up interview.}
\label{tab:follow}
\Description{Follow-up interview questions cover four domains: usage frequency; feature/mode utilization (patterns, tasks, benefits, limitations); usage experience and perceived impact (understanding, review process, impression changes, insights); and challenges plus desired feature improvements.}
\resizebox{\textwidth}{!}{
\begin{tabular}{p{0.15\textwidth}p{0.8\textwidth}}
\toprule
\textbf{Category} & \textbf{Question Phrase} \\
\midrule
Frequency of Use & $\cdot$ How often have you used Copilot in the past two weeks? Why? \\
\hline
\multirow{3}{*}{Modes} & $\cdot$ Which modes or features did you use more frequently during these two weeks? \\
 & $\cdot$ What specific tasks did you use them for? \\
 & $\cdot$ What benefits or shortcomings did you notice in their usage? \\
\hline
\multirow{6}{*}{\makecell[l]{Usage Experience \\ \& Impact}} & $\cdot$ Do you think Copilot understands your needs well? \\
& $\cdot$ How do you prefer to review responses now?  \\
& $\cdot$ Do you find the process of reviewing and accepting responses smooth? Were there any obstacles or inconveniences? \\
 & $\cdot$ After continuous use of Copilot, has your overall impression of the tool changed? What new insights or discoveries have you made? \\
\hline
\multirow{3}{*}{\makecell[l]{Challenges \\ \& Feedback}}& $\cdot$ During the past two weeks, did you encounter any new difficulties or challenges while using Copilot? \\
 & $\cdot$ After using Copilot for a while, what features or improvements would you most like to see? \\
\bottomrule
\end{tabular}
}
\end{table}

\section*{Appendix 3: All Users' Interaction Timelines}
\begin{figure}[h]
    \centering
    \includegraphics[height=\textheight]{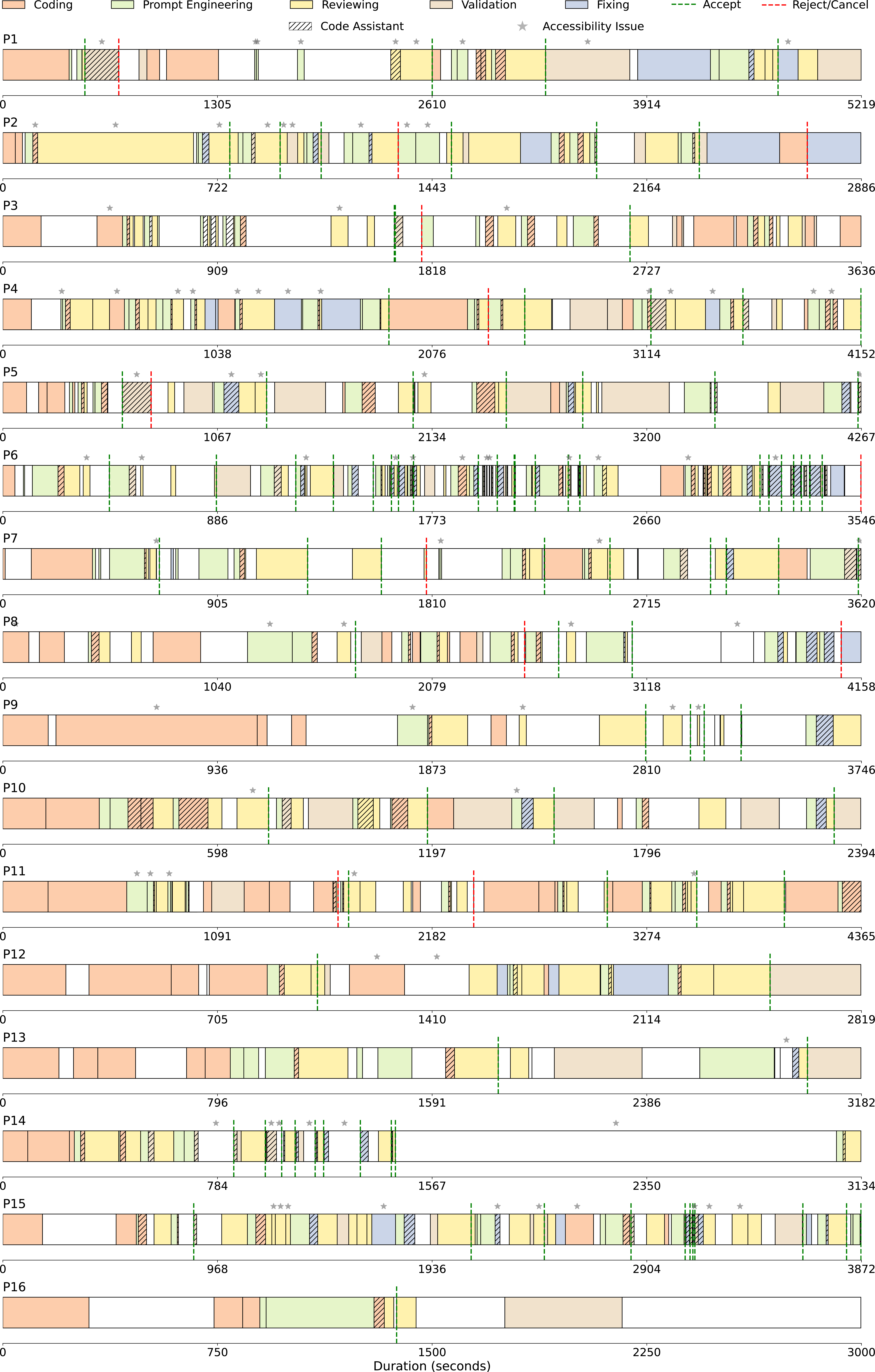}
    \vspace{-1em}
    \caption{All users' action timelines during their programming task.}
    \label{fig:all_timelines}
    \vspace{-1em}
    \Description{All users' action timelines highlight varied, human-dominated workflows with frequent review cycles.
    Sixteen horizontal tracks of user activities (prompting, reviewing, validation, coding, fixing, waiting) interspersed with brief assistant action markers. Overall, assistant actions are short and consistently bracketed by human oversight.}
\end{figure}

Finally, we presented all 16 users' interaction timelines during their programming task in \autoref{fig:all_timelines}.

\end{document}